%% file: main.tex
\documentclass{article}

\usepackage{arxiv}

\input{0.mypackages}

\input{0.metadata}

\begin{document}

\input{0.acronyms}

\maketitle

\begin{abstract}
\input{0.abstract}
\end{abstract}


\glsresetall

\input{_sec_text}

\input{0.ack}

\bibliographystyle{plainnat}
\bibliography{0.sample-base}

\end{document}

%% file: 0.mypackages.tex
\newcommand{\code}[1]{\texttt{\small #1}}

\usepackage[utf8]{inputenc}
\usepackage[T1]{fontenc}
\usepackage{microtype}

\usepackage[table,dvipsnames]{xcolor}

\usepackage{array}                     
\usepackage{booktabs}                  
\usepackage{tabularx}
\usepackage{longtable}
\usepackage{multirow}
\usepackage{makecell}
\usepackage{rotating}
\usepackage{colortbl}                  

\newcolumntype{L}[1]{>{\raggedright\arraybackslash}p{#1}}
\newcolumntype{C}[1]{>{\centering\arraybackslash}p{#1}}

\usepackage{enumitem}
\usepackage{siunitx}
\usepackage{comment}
\usepackage{setspace}
\usepackage{url}                       

\usepackage{amsmath}                   
\usepackage{amssymb}                   
\usepackage{amsfonts}                  
\usepackage{amsthm}                    
\usepackage{nicefrac}                  

\newtheorem{lemma}{Lemma}
\theoremstyle{definition}

\AtBeginDocument{%
  }
\providecommand{\kesbmc}{\textsc{\mbox{K-ESBMC}}}
\providecommand{\kst}{\textsc{\mbox{K-ST}}}
\providecommand{\esbmcplc}{\textsc{\mbox{ESBMC-PLC}}}
\providecommand{\openplc}{OpenPLC}
\providecommand{\cmark}{\ensuremath{\checkmark}}      
\providecommand{\xmark}{\ensuremath{\boldsymbol{\times}}}
\providecommand{\Description}[1]{}

\usepackage{graphicx}
\usepackage{tikz}
\usetikzlibrary{
    positioning,
    shapes.geometric,
    shapes.misc,
    arrows.meta,
    calc,
    mindmap,
    shadows,
    fit,
    backgrounds,
    patterns
}
\usepackage{pgfplots}
\pgfplotsset{compat=1.18}

\usepackage{algorithm}
\usepackage{algpseudocode}

\definecolor{amber}{rgb}{0.75, 0.50, 0.00}
\definecolor{teal} {rgb}{0.00, 0.50, 0.50}

\definecolor{reviewerblue} {RGB}{0,70,140}
\definecolor{commentgray}  {RGB}{240,240,245}
\definecolor{responsegreen}{RGB}{0,100,60}

\definecolor{grpFormal}{RGB}{180,160,220}   
\definecolor{grpPlan}  {RGB}{140,210,185}   
\definecolor{grpAgent} {RGB}{250,185,130}   
\definecolor{grpRun}   {RGB}{250,215,100}   
\definecolor{grpBench} {RGB}{195,195,185}   

\definecolor{ldblue}  {HTML}{1565C0}
\definecolor{trgreen} {HTML}{1B5E20}
\definecolor{cpurp}   {HTML}{4A148C}
\definecolor{esorg}   {HTML}{BF360C}
\definecolor{safegrn} {HTML}{2E7D32}
\definecolor{unsafrd} {HTML}{B71C1C}
\definecolor{cexbrn}  {HTML}{4E342E}
\definecolor{diagpnk} {HTML}{880E4F}
\definecolor{panelbg} {HTML}{F5F5F5}

\definecolor{figviolet}{RGB}{180,160,220}   
\definecolor{figamber} {RGB}{250,185,100}   
\definecolor{figteal}  {RGB}{180,225,210}   
\definecolor{figblue}  {RGB}{180,210,240}   
\definecolor{figgray}  {RGB}{220,220,215}   

\definecolor{codebg}    {HTML}{F8F8F8}
\definecolor{codeframe} {HTML}{DDDDDD}
\definecolor{codegreen} {HTML}{2E7D32}
\definecolor{codeblue}  {HTML}{1565C0}
\definecolor{codegray}  {HTML}{757575}
\definecolor{codepurple}{HTML}{7B1FA2}

\usepackage{listings}
\usepackage{eurosym}               

\newcommand{\bars}[1]{%
  \begin{tikzpicture}[baseline=-0.4ex]
    \foreach \i in {1,...,5}{%
      \pgfmathparse{\i <= #1 ? 1 : 0}%
      \ifnum\pgfmathresult=1
        \fill[black]     (\i*0.12-0.12, 0) rectangle (\i*0.12-0.03, 0.22);
      \else
        \fill[black!18]  (\i*0.12-0.12, 0) rectangle (\i*0.12-0.03, 0.22);
      \fi
    }
  \end{tikzpicture}%
}

\usepackage[numbers]{natbib}

\usepackage{hyperref}

\usepackage[acronym, nogroupskip, nonumberlist, nopostdot]{glossaries}
\glsdisablehyper

%% file: 0.metadata.tex
\title{\kesbmc{}: An Executable Formal Semantics of IEC~61131-3 Ladder Diagram for Validating Verifier Translations}

\renewcommand{\shorttitle}{\textsc{K-ESBMC}: Validating a Ladder-Diagram Verifier's Front-End}

\author{
  Pierre Dantas\thanks{The authors contributed equally to this research.}\\
  Department of Computer Science\\
  The University of Manchester\\
  Manchester, UK\\
  \texttt{pierre.dantas@manchester.ac.uk}\\
  ORCID: 0000-0001-6390-9340\\
  \And
  Lucas Cordeiro\footnotemark[1]\\
  Department of Computer Science\\
  The University of Manchester\\
  Manchester, UK\\
  \texttt{lucas.cordeiro@manchester.ac.uk}\\
  ORCID: 0000-0002-6235-4272\\
  \And
  Waldir Junior\footnotemark[1]\\
  Electrical Engineering\\
  Federal University of Amazonas (UFAM)\\
  Manaus, AM, Brazil\\
  \texttt{waldirjr@ufam.edu.br}\\
  ORCID: 0000-0003-3095-0042\\
}

\date{July 2026}

%% file: 0.acronyms.tex
\newacronym{aadl}{AADL}{Architecture Analysis and Design Language}
\newacronym{ansi-c}{ANSI-C}{American National Standards Institute C}
\newacronym{api}{API}{Application Programming Interface}
\newacronym{adc}{ADC}{Analog-to-Digital Converter}
\newacronym{ast}{AST}{Abstract Syntax Tree}
\newacronym{bdd}{BDD}{Binary Decision Diagrams}
\newacronym{bmc}{BMC}{Bounded Model Checking}
\newacronym{cbmc}{CBMC}{Bounded Model Checking for ANSI-C Programs}
\newacronym{cegar}{CEGAR}{Counterexample-Guided Abstraction Refinement}
\newacronym{cern}{CERN}{Conseil Européen pour la Recherche Nucléaire}
\newacronym{cfg}{CFG}{Control Flow Graph}
\newacronym{chc}{CHC}{Constrained Horn Clause}
\newacronym{cli}{CLI}{Command-Line Interface}
\newacronym{cpu}{CPU}{Central Processing Unit}
\newacronym{ctl}{CTL}{Computation Tree Logic}
\newacronym{cuda}{CUDA}{Compute Unified Device Architecture}
\newacronym{cve}{CVE}{Common Vulnerability and Exposure}
\newacronym{dfs}{DFS}{Depth-First Search}
\newacronym{dsl}{DSL}{Domain-Specific Language}
\newacronym{epsrc}{EPSRC}{Engineering and Physical Sciences Research Council}
\newacronym{evm}{EVM}{Ethereum Virtual Machine}
\newacronym{esbmc}{ESBMC}{Efficient SMT-based Context-Bounded Model Checker}
\newacronym{fbd}{FBD}{Functional Block Diagram}
\newacronym{fpga}{FPGA}{Field-Programmable Gate Array}
\newacronym{gpio}{GPIO}{General-Purpose Input/Output}
\newacronym{hal}{HAL}{Hardware Abstraction Layer}
\newacronym{hil}{HIL}{Hardware-in-the-Loop}
\newacronym{ic3}{IC3}{Incremental Construction of Inductive Clauses for Indubitable Correctness}
\newacronym{ide}{IDE}{Integrated Development Environment}
\newacronym{iec}{IEC}{International Electrotechnical Commission}
\newacronym{ieee}{IEEE}{Institute of Electrical and Electronics Engineers}
\newacronym{ics}{ICS}{Industrial Control Systems}
\newacronym{il}{IL}{Instruction List}
\newacronym{iot}{IoT}{Internet of Things}
\newacronym{ir}{IR}{Intermediate Representation}
\newacronym{iso}{ISO}{International Organization for Standardization}
\newacronym{ld}{LD}{Ladder Diagram}
\newacronym{llb}{LLB}{Ladder Logic Bombs}
\newacronym{llm}{LLM}{Large Language Model}
\newacronym{ltl}{LTL}{Linear Temporal Logic}
\newacronym{matiec}{MATIEC}{IEC 61131-3 compiler}
\newacronym{mcu}{MCU}{Microcontroller Unit}
\newacronym{nasa}{NASA}{National Aeronautics and Space Administration}
\newacronym{pwm}{PWM}{Pulse-Width Modulation}
\newacronym{pdr}{PDR}{Property Directed Reachability}
\newacronym{pid}{PID}{Proportional-Integral-Derivative}
\newacronym{plc}{PLC}{Programmable Logic Controller}
\newacronym{pou}{POU}{Program Organization Unit}
\newacronym{por}{POR}{Partial Order Reduction}
\newacronym{rtos}{RTOS}{Real-Time Operating System}
\newacronym{sat}{SAT}{Boolean Satisfiability}
\newacronym{scl}{SCL}{Structured Control Language}
\newacronym{sfc}{SFC}{Sequential Function Chart}
\newacronym{slr}{SLR}{Systematic Literature Review}
\newacronym{smt}{SMT}{Satisfiability Modulo Theories}
\newacronym{smtlib2}{SMT-LIB}{Satisfiability Modulo Theories Library}
\newacronym{ssa}{SSA}{Static Single Assignment}
\newacronym{st}{ST}{Structured Text}
\newacronym{stl}{STL}{Statement List}
\newacronym{svcomp}{SV-COMP}{Competition on Software Verification}
\newacronym{tacas}{TACAS}{Tools and Algorithms for the Construction and Analysis of Systems}
\newacronym{ufam}{UFAM}{Federal University of Amazonas}
\newacronym{ukri}{UKRI}{UK Research and Innovation}

%% file: 0.abstract.tex
Automated verifiers for \acrshort{iec}~61131-3 \gls{ld} enhance safety by translating diagrams into model-checker inputs. Still, their unverified front-end translations risk silently returning incorrect results (missing violations or raising false alarms) when they diverge from the standard. We address this gap with \textbf{\kesbmc{}}, an executable formal semantics of \acrshort{iec}~61131-3 \gls{ld} built in the K framework. \kesbmc{} models contacts, coils, timers, counters, edge blocks, and the retentive scan cycle, generating both an interpreter and a deductive verifier from a single definition. Validated scan-for-scan against \openplc{}/\acrshort{matiec}; \kesbmc{} serves as an independent reference oracle to test the \esbmcplc{} \gls{ld}$\rightarrow$GOTO translation differentially. It agrees with \acrshort{esbmc} on most programs and reproduces injected violations with concrete witnesses. Every disagreement exposes a genuine \acrshort{esbmc} defect -- confirmed by \openplc{} and two other verifiers -- revealing two failure modes: an unsound \emph{skip} that certifies unsafe programs, and an imprecise \emph{havoc} that produces spurious counterexamples. For the combinational and latch fragment, we machine-check in \texttt{kprove} that \kesbmc{}'s rules implement the standard's input/output relation, elevating the correctness argument from empirical to formal. \kesbmc{} provides a reusable, standard-faithful oracle for auditing any \gls{ld} verifier's translation, offering a general approach to verifying the soundness of translation-based verification tools.

%% file: _sec_text.tex
\section{Introduction}
\label{sec:intro}

\Glspl{plc} run much of the world's safety-critical automation, and \gls{ld} -- the graphical, relay-inspired notation of \gls{iec}~\mbox{61131-3} -- remains the language in which that logic is most often written and certified. Because a \gls{plc} fault can injure people or damage a plant, \gls{ld} programs are natural targets for formal verification. A line of automated tools now checks safety properties of \gls{iec}~\mbox{61131-3} programs by translating them into the input language of a model checker. The \esbmcplc{}  family~\cite{DantasCordeiro2026artefact, DantasCordeiro2026graphical, ESBMCpr5400}, for instance, lowers an \gls{ld} diagram into a GOTO \gls{ir} and discharges the resulting verification conditions with \gls{smt}-based \gls{bmc} and $k$-induction.

\subsection{The Trust Gap} 
Every such tool rests on an unstated assumption: that its \emph{front-end} -- the translation from the diagram to the verifier's model --  faithfully reflects \gls{iec}~\mbox{61131-3}. A verifier is only as sound as this translation, yet the translation is itself unverified code. In the \esbmcplc{} line, the \gls{ld}$\rightarrow$GOTO rules were \emph{designed to match} an informal reference formalization, but were never proven equivalent to the standard. When the translation and the standard disagree, the verifier can silently return the wrong answer: it can certify a hazardous program as safe (if the translation drops behavior that the standard mandates) or raise false alarms (if it admits behavior that the standard forbids). Neither failure is detectable from within the tool, because there is no independent, executable account of what the diagram \emph{means}.

\subsection{An Executable Reference Semantics} 
We close this gap with \kesbmc{}, an executable formal semantics of \gls{iec}~\mbox{61131-3} \gls{ld} defined in the K framework. K lets one write an operational semantics as a set of rewrite rules and obtain, for free, both an interpreter (\code{krun}) and a program verifier (\code{kprove}) from the same definition. \kesbmc{} extends \kst{}~\cite{Wang2023}, the K semantics of \gls{iec}~\mbox{61131-3} \gls{st}, to the graphical \gls{ld} language: it models normally-open and normally-closed contacts, energize/latch/unlatch coils, the input-solve-output \emph{scan cycle} with cross-scan state retention, the on-/off-delay and pulse timers (\code{TON}, \code{TOF}, \code{TP}), the up/down counters (\code{CTU}, \code{CTD}), and the edge-detection function blocks (\code{R\_TRIG}, \code{F\_TRIG}). Being executable, \kesbmc{} is not a paper artifact: it \emph{runs} diagrams scan by scan, so its fidelity can be checked directly against a reference \gls{plc} runtime (Fig.~\ref{fig:kesbmc-overview}).

\input{fig_kesbmc_overview}

\subsection{Grounding and Using the Semantics} 
We confirm \kesbmc{}'s fidelity and then put it to work. First, we validate \kesbmc{} scan-for-scan against the reference function-block implementations that \openplc{} executes -- the \gls{matiec}-compiled C that is the de~facto behavioral standard for these constructs -- so that \kesbmc{}'s timers and counters agree with real \gls{plc} execution on every scan. Second, we use \kesbmc{} as an \emph{independent reference oracle} to validate the \esbmcplc{} differentially \gls{ld}$\rightarrow$GOTO translation on a suite of benchmark controllers. For each program, we compare the verifier's verdict against the ground truth obtained by executing the diagram under \kesbmc{}. Third, for the combinational and latch fragments, we go beyond execution and \emph{machine-check} in \code{kprove} that \kesbmc{}'s rules implement the \gls{iec}~\mbox{61131-3} input/output relation for each construct. This combination makes the reference semantics explicit, executable, and independently checkable.

\subsection{Why These Defects Matter} 
An \gls{ld} program is not free-standing software: it is the controller of a physical process, executed on the cyclic \emph{scan} that couples computation to physical time. The constructs on which we find defects -- the timers -- are precisely the ones that encode that coupling, turning elapsed physical time into control decisions (an off-delay that keeps a light on, an on-delay that sequences a traffic phase, a debounce that gates an actuator). A translation error in a timer is therefore not a cosmetic issue, yet a misprediction of the controller's real behavior with safety consequences: in one of our benchmarks, the verifier certifies as safe a stairwell controller whose light, driven by an off-delay, stays on while the corridor is empty, because the front-end drops the timer and still reports the program verified. Catching such a defect requires a semantics that models the scan cycle and function-block timing faithfully enough to reproduce what the controller actually does, and validating that semantics requires checking it against the runtime deployed on real hardware. That is what \kesbmc{} provides.

The differential study is revealing. On the benchmark suite, \kesbmc{} and \gls{esbmc} agree on the vast majority of programs. For the unsafe variants, \kesbmc{} independently reproduces every violation, localizing it to the offending property with a concrete witness. Every \emph{disagreement}, however, is a genuine \gls{esbmc} defect that \kesbmc{} exposes and \openplc{} confirms -- and they fall into two opposite classes, both rooted in \gls{esbmc}'s incomplete handling of timer function blocks. On graphical diagrams \gls{esbmc} \emph{skips} the timer path and thereby \emph{misses a real violation}, certifying as safe a program that violates its own stated property (an unsound, dangerous error). In the simple diagram format, it instead leaves the timer outputs \emph{unconstrained}, yielding counterexamples that no real timer can produce (a false alarm that wastes engineering effort). An executable, standard-faithful semantics is exactly the instrument needed to tell these apart from true defects.

\subsection{Contributions}
\label{sub:contributions}
This paper yields:
\begin{enumerate}
    \item \textbf{\kesbmc{}, an executable formal semantics of \gls{iec}~\mbox{61131-3} \gls{ld} in K} (\S\ref{sec:semantics}), extending \kst{} to the graphical language: contacts, coils and latches, the retentive scan cycle, timers (\code{TON}/\code{TOF}/\code{TP}), counters (\code{CTU}/\code{CTD}), and edge function blocks. From one definition, we obtain both an interpreter and a deductive verifier.
    
    \item \textbf{Fidelity comparison against \openplc{}/\gls{matiec}}: \kesbmc{}'s function-block semantics are shown to agree scan-for-scan with the reference implementations that \openplc{} executes, grounding \kesbmc{} as a behavioral reference for real \glspl{plc}.
    
    \item \textbf{A differential validation of the \esbmcplc{} \gls{ld}$\rightarrow$GOTO translation} using \kesbmc{} as an independent oracle (\S\ref{sec:e3}), over a benchmark suite covering the full construct fragment and both simple and graphical diagram formats. We report broad agreement and three discrepancies -- each corroborated by the \openplc{} runtime and, on the timer mechanism they share, by two further independent verification engines (\gls{matiec}$\to$C$+$\gls{cbmc} and NuSMV) -- and classify them into \gls{esbmc}'s two timer-translation failure modes: unsound \emph{skip} (missed bugs) and imprecise \emph{havoc} (false alarms).
    
    \item \textbf{Machine-checked construct-correctness lemmas} for the combinational and latch fragment, discharged in \code{kprove} (\S\ref{sec:eval:rq4}), raising the correctness argument for that fragment from empirical to formally proven.
\end{enumerate}

\subsection{Roadmap}
\S\ref{sec:background} reviews \gls{ld} and the scan cycle, the K framework and \kst{}, the \openplc{}/\gls{matiec} toolchain, and the \esbmcplc{} translation, and \S\ref{sec:related} surveys related work. \S\ref{sec:semantics} presents \kesbmc{}, its testing against \openplc{} (RQ1), and the mechanized construct lemmas (RQ4); \S\ref{sec:e3} reports the differential study (RQ2/RQ3); \S\ref{sec:cases} works through two case studies; and \S\ref{sec:discussion}--\S\ref{sec:conclusion} discuss implications, limitations, and conclusions.

\subsection{Notation and Conventions}
\label{sub:notation}
Table~\ref{tab:notation} collects the notation used throughout. For \gls{ld} elements we use the familiar contact/coil mnemonics (\code{XIC}, \code{XIO}, \code{OTE}, \code{OTL}, \code{OTU}) and the \gls{iec}~\mbox{61131-3} standard function-block names (\code{TON}, \code{TOF}, \code{TP}, \code{CTU}, \code{CTD}, \code{R\_TRIG}, \code{F\_TRIG}). Boolean signal values are written $\top$ (true / energized) and $\bot$ (false / de-energized). The symbols $\Rightarrow$ and $\rightsquigarrow$ denote K's rewrite and sequencing relations, and $\langle\code{c}\rangle$ denotes a configuration cell~\code{c} of the semantics.

\begin{table}[t]
    \centering
    \caption{Notation and conventions used throughout the paper.}
    \label{tab:notation}
    \small
    \begin{tabular}{@{}l@{\hspace{1.5em}}p{0.66\linewidth}@{}}
        \toprule
        Symbol & Meaning \\
        \midrule
        \multicolumn{2}{@{}l}{\emph{Ladder elements (contacts and coils)}} \\\midrule
        \code{XIC} & normally-open contact -- passes power when its variable is $\top$ \\
        \code{XIO} & normally-closed contact -- passes power when its variable is $\bot$ \\
        \code{OTE} & energize coil -- copies the rung condition to its variable \\
        \code{OTL}, \code{OTU} & latch / unlatch coils -- set / reset, retained across scans \\
        \addlinespace
        \multicolumn{2}{@{}l}{\emph{Function blocks}} \\\midrule
        \code{TON}, \code{TOF}, \code{TP} & on-delay, off-delay, and pulse timers \\
        \code{CTU}, \code{CTD} & up and down counters \\
        \code{R\_TRIG}, \code{F\_TRIG} & rising- and falling-edge detectors \\
        \addlinespace
        \multicolumn{2}{@{}l}{\emph{Values and logic}} \\\midrule
        $\top$, $\bot$ & Boolean true / false (energized / de-energized) \\
        $\lnot$, $\land$, $\lor$ & negation; series (conjunction) and parallel (disjunction) of contacts \\
        \addlinespace
        \multicolumn{2}{@{}l}{\emph{Timing and block parameters}} \\\midrule
        $\Delta t$ & scan cycle time -- physical time elapsed per scan \\
        $\mathrm{PT}$, $\mathrm{ET}$ & timer -- Preset Time and Elapsed Time \\
        $\mathrm{PV}$, $\mathrm{CV}$ & counter - Preset Value and Current Value \\
        $\lceil \mathrm{PT}/\Delta t\rceil$ & number of scans after which a timer fires \\
        \addlinespace
        \multicolumn{2}{@{}l}{\emph{Semantics (K framework)}} \\\midrule
        $\langle\code{c}\rangle$ & a configuration cell (e.g.\ $\langle\code{image}\rangle$, $\langle\code{timers}\rangle$, $\langle\code{dt}\rangle$) \\
        $\code{image}[X]$ & current Boolean value of signal $X$ in the retained image \\
        $\Rightarrow$ & K rewrite -- one term reduces to another \\
        $\rightsquigarrow$ & K sequencing of computations \\
        $\gls{ld}\rightarrow$GOTO & the \gls{esbmc} front-end translation under validation \\
        \bottomrule
    \end{tabular}
\end{table}

\section{Background}
\label{sec:background}

\subsection{\Acrfull{ld} and the Scan Cycle}
\label{sub:ld}
An \gls{iec}~\mbox{61131-3} \gls{ld} program is a ladder of \emph{rungs} drawn between two power rails. Along a rung, \emph{contacts} test Boolean signals -- a normally-open contact (\code{XIC}) passes power when its variable is true, a normally-closed contact (\code{XIO}) when it is false -- and series and parallel arrangements of contacts realize conjunction and disjunction. Power reaching the right rail drives a \emph{coil}: an energize coil (\code{OTE}) copies the rung condition to its variable, while latch and unlatch coils (\code{OTL}/\code{OTU}) set or reset it and \emph{retain} the value across scans. Rungs may also contain \emph{function-block} instances -- timers (\code{TON}, \code{TOF}, \code{TP}), counters (\code{CTU}, \code{CTD}), and edge detectors (\code{R\_TRIG}, \code{F\_TRIG}) -- each carrying private state.

A \gls{plc} executes a program by the cyclic \emph{scan}: it samples the inputs, solves every rung top to bottom, writes the outputs, and repeats. State that is not recomputed each scan -- latched coils, timer accumulators, counter values --  persists between scans, which is precisely what makes timing and sequencing behavior possible and what a faithful semantics must model. Diagrams are exchanged in the \textbf{PLCopen XML} format, which encodes each rung as a netlist of contact, coil, and block elements wired by identifier references; the benchmarks used here are PLCopen XML documents.

\subsection{System and Timing Model}
\label{sub:sysmodel}
The system we reason about is a \gls{plc} in closed interaction with a physical plant: each scan reads sensor inputs, computes the control logic, and writes actuator outputs, and the plant evolves in the interval before the next scan. We adopt the standard synchronous \emph{scan-cycle} abstraction of this cyber-physical loop. Inputs are sampled once per scan and held constant during the solve; outputs take effect at the end of the scan; and physical time is discretized into scans of uniform cycle time~$\Delta t$, so that a function block's elapsed time advances by exactly $\Delta t$ per scan. Under this model, a timer with preset~$\mathrm{PT}$ acts after $\lceil\mathrm{PT}/\Delta t\rceil$ scans -- the point at which computation and physical timing meet, and, as \S\ref{sec:e3} shows, the point at which a verifier's front-end most easily departs from the physical behavior of the controller.

Two assumptions delimit the scope. First, we model the \emph{control-logic} layer at the granularity exposed by the standard and \gls{plc} runtime -- Boolean signals, retained function-block state, and the scan schedule -- and abstract continuous dynamics and sub-scan jitter, as standard for scan-cycle verification. Safety properties are consequently over the controller's I/O behavior per scan (e.g., ``the light is on only while the sensor is active''), which is where the discrepancies of \S\ref{sec:e3} live. Second, we assume no faults or adversary: the failures we expose are \emph{translation} defects in the verifier, not plant faults or attacks. This timing model is exactly what \kesbmc{} implements (\S\ref{sec:semantics}) and the reference against which we validate it using the runtime deployed on real open hardware (\S\ref{sec:eval:rq1}).

\subsection{The K Framework and \kst{}}
\label{sub:kframework}

The K framework~\cite{Rosu2010k} is a semantic framework in which the formal semantics of a language is given as a configuration (a structured collection of cells) together with rewrite rules over it. From a single definition, K generates an executable interpreter and a reachability-logic prover (\code{kprove}), so the \emph{same} artifact both runs programs and proves properties about them -- an excellent match for a reference semantics that must be simultaneously trustworthy and checkable. K has been used to give complete executable semantics to C, Java, JavaScript, and the \gls{evm}, among others. Closest to our work, \citet{Wang2023} define \kst{}, a K semantics for \gls{iec}~\mbox{61131-3} \gls{st}. \kesbmc{} adopts the same framework and scan-cycle discipline but targets the graphical \gls{ld} language and its function blocks, which \kst{} does not cover (Fig.~\ref{fig:k-framework}).

\input{fig_k_framework}

\subsection{\openplc{}, \gls{matiec}, and Reference Behavior}
\label{sub:openplc}
\openplc{} is a widely used open-source \gls{plc} runtime; it compiles \gls{iec}~\mbox{61131-3} programs to C using \gls{matiec}~\cite{deSousa2014} and executes them on a scan loop. Crucially, \gls{matiec} ships reference C implementations of the standard function blocks -- the \code{TON}, \code{TOF}, \code{TP}, \code{CTU}, and \code{CTD} bodies -- driven by a controlled notion of elapsed time. Because these implementations are what actually run on deployed open-hardware \glspl{plc}, we treat their per-scan behavior as the behavioral reference against which \kesbmc{}'s function blocks are validated, and as the tie-breaker whenever a verifier and \kesbmc{} disagree (Fig.~\ref{fig:matiec-reference}).

\input{fig_matiec_reference}

\subsection{Verifying \gls{ld}: \esbmcplc{} and the Translation Gap}
\label{sub:esbmcplc}
\gls{esbmc}~\cite{gadelha2020} is a mature \gls{smt}-based \gls{bmc} and $k$-induction tool for C and other languages. The \esbmcplc{} line adds an \gls{ld} front-end that parses PLCopen XML, lowers each rung into a GOTO \gls{ir} modeling one scan, wraps it in a scan loop, and checks user-supplied safety properties. The lowering is designed to agree with an informal reference formalization of \gls{ld}~\cite{Ebnenasir2023}, but there is no proof that it preserves \gls{iec}~\mbox{61131-3} semantics. This is the gap \kesbmc{} addresses: it supplies the missing executable and a standard-faithful account of \gls{ld} against which the lowering can be checked -- empirically via differential testing and partially via machine-checked lemmas.

\section{Related Work}
\label{sec:related}

\subsection{Formal Semantics of \gls{iec}~\mbox{61131-3}} 
Formal treatments of \gls{iec}~\mbox{61131-3} date back two decades~\cite{Frey2000formal}, using timed automata, transition systems, and abstract state machines to enable model checking of \gls{st}, \gls{il}~\cite{Canet2000il}, or \glspl{sfc}. These semantics are typically defined \emph{on paper} or embedded in a specific model checker, and serve a single tool. \kst{}~\cite{Wang2023} is the first K semantics for the family and the direct ancestor of this work; \kesbmc{} extends the framework to the graphical \gls{ld} language and its function blocks. Unlike semantics built solely to feed a model checker, an executable K definition doubles as a reference interpreter, which is what makes the differential validation of another tool possible.

\subsection{\gls{plc} Verification Tools} 
A range of tools verify \gls{iec}~\mbox{61131-3} controllers -- among them PLCverif~\cite{Darvas2015plcverif,Darvas2014forte}, Arcade.PLC~\cite{Biallas2012arcade}, and SMV/nuXmv-based flows~\cite{Cavada2014}, some scaled to industrial-sized programs~\cite{FernandezAdiego2015tii} -- each translating the program into a model checker's input. Table~\ref{tab:related} contrasts them with \kesbmc{} along the axes that matter here. All of these tools are \emph{verifiers}: they consume a diagram and emit a verdict; each therefore embeds a source-to-model front-end whose fidelity is assumed rather than verified. None is an executable reference semantics, and none validates its own translation against an independent account of \gls{ld}. \kesbmc{} is deliberately \emph{not} another verifier: it is the reference oracle against which such a translation can be scrutinized, and we apply it to \esbmcplc{}.\footnote{We apply the differential to \esbmcplc{} rather than to these other tools because none can ingest our PLCopen~XML \gls{ld} benchmarks without a further trusted translation: PLCverif's only \gls{plc} front-end targets Siemens Step7 (\acrshort{scl}/STL), and Arcade.PLC is no longer distributed. Instead, we support the disagreements using verification engines whose decision procedures are independent of \kesbmc{} (\S\ref{sec:e3:rq3}).} Unlike PLCverif and Arcade.PLC, which is tied to specific backends, \kesbmc{} is back-end-agnostic and, being executable, is validated against the runtime (\openplc{}) on which the deployed controllers actually run.

\begin{table}[t]
    \centering
    \caption{\kesbmc{} versus representative \gls{iec}~\mbox{61131-3} semantics and verification tools. ``Executable'' = runs programs to produce concrete traces; ``Oracle'' = usable to validate another tool's translation; ``Runtime-validated'' = checked against a deployed \gls{plc} runtime.}
    \label{tab:related}
    \small
    \begin{tabular}{lccccc}
        \toprule
        & \gls{ld} & Function & Execu- & Reference & Runtime- \\
        & graphical & blocks & table & oracle & validated \\
        \midrule
        PLCverif        & partial & yes & no  & no  & no  \\
        Arcade.PLC      & yes     & yes & no  & no  & no  \\
        SMV/nuXmv flows & partial & yes & no  & no  & no  \\
        \kst{}~\cite{Wang2023} (ST) & n/a & no & yes & (unused) & no \\
        \esbmcplc{} family  & yes & partial & no & no & no \\
        \midrule
        \textbf{\kesbmc{}} & \textbf{yes} & \textbf{yes} & \textbf{yes} & \textbf{yes} & \textbf{yes} \\
        \bottomrule
    \end{tabular}
\end{table}

\subsection{Translation Validation, Differential Testing, and Semantics in K}
Establishing that a front-end or compiler preserves semantics is the subject of translation validation~\cite{Pnueli1998translation} and verified compilation~\cite{Leroy2009compcert}. Full verified compilation of a \gls{plc} front-end would require a automated semantics of both the source diagram and the verifier's \gls{ir}; we take the lighter, high-coverage route of \emph{differential testing}~\cite{Yang2011csmith} against an executable reference -- the same principle by which compiler bugs are found by comparing implementations -- complemented by machine-checked per-construct lemmas. Using an executable K semantics for both testing and proving is enabled by K's dual interpreter/prover generation. It parallels how complete K semantics of mainstream languages -- C~\cite{Ellison2012c} and the \gls{evm}~\cite{Hildenbrandt2018kevm} among them -- have been used to find and rule out discrepancies in real implementations. What is new here is turning that reference-semantics methodology on a \emph{verifier's} front-end in the cyber-physical \gls{plc} setting, where the discrepancies remain safety-relevant.

\input{fig_positioning}

\section{The \kesbmc{} Semantics and Its Validation}
\label{sec:semantics}

Our evaluation follows one principle: an oracle must be shown trustworthy \emph{before} it is used to judge another tool, and its judgment is most convincing when it is grounded in real execution and, where tractable, proven. This principle fixes four research questions and the order in which they must be answered, each with a specific purpose:

\begin{itemize}

    \item\textbf{RQ1} (\S\ref{sec:eval:rq1}) -- is \kesbmc{} faithful to \gls{iec}~\mbox{61131-3}? --  \emph{grounds the instrument}: we validate \kesbmc{} scan-for-scan against the code \openplc{} executes, because any verdict from an unfaithful oracle would be worthless. 
    
    \item\textbf{RQ2} (\S\ref{sec:e3:rq2}) -- does \kesbmc{}, used as an oracle, agree with the \gls{esbmc} \gls{ld}$\rightarrow$GOTO translation on programs both handle? -- tests the oracle's \emph{credibility} as an independent second opinion. 
    
    \item\textbf{RQ3} (\S\ref{sec:e3:rq3}) -- where they disagree, which engine is correct and why? -- exercises the oracle's \emph{diagnostic power}, the payoff of the whole exercise, by adjudicating each disagreement against real execution and classifying the translation defects it exposes. 
    
    \item\textbf{RQ4} (\S\ref{sec:eval:rq4}) -- can \kesbmc{}'s construct rules be \emph{machine-checked}? -- \emph{hardens the foundation}, so the oracle rests on proof, not only on execution, wherever the proof is tractable. 
    
\end{itemize}

In short, RQ1 grounds the instrument, RQ2 and RQ3 apply it, and RQ4 certifies it; RQ1 and RQ4 are answered in this section, RQ2 and RQ3 in the differential study of \S\ref{sec:e3}.

\subsection{The Semantics in \texorpdfstring{K}{K}}
\label{sub:sem}
\kesbmc{} is a configuration of cells rewritten scan by scan. The retained state is a single Boolean \emph{image} ($\langle\code{image}\rangle$) mapping every input, coil, and function-block done-bit to its current value; alongside it, $\langle\code{timers}\rangle$ and $\langle\code{counters}\rangle$ hold per-instance function-block state, and $\langle\code{dt}\rangle$ is the cycle time~$\Delta t$. The driver cell $\langle\code{k}\rangle$ runs the scan loop over an input trace $\langle\code{inputs}\rangle$ (one Boolean assignment per scan), logging the image after each scan to $\langle\code{trace}\rangle$. One scan overlays the next input, solves the program's rungs top to bottom against the image, and appends the result:
\[
\text{\code{\#run}} \Rightarrow P \,\rightsquigarrow\, \text{\code{\#endScan}}
\,\rightsquigarrow\, \text{\code{\#run}}
\qquad\text{(consuming one input, per scan).}
\]
Contacts read the image ($\code{XIC}(X)\Rightarrow \code{image}[X]$; $\code{XIO}(X)\Rightarrow\lnot\,\code{image}[X]$), series and parallel compose by conjunction and disjunction, and \code{OTE}/\code{OTL}/\code{OTU} write, set, or retain their coil. Because only recomputed coils change each scan while latches and function-block state persist in their cells, cross-scan retention -- the crux of timing and sequencing -- is modeled directly. The timers are a faithful port of the standard state machine: each carries $(\mathrm{ET},\mathrm{STATE},\mathrm{prevIN})$ with $\mathrm{STATE}\in\{\text{idle},\text{timing},\text{done}\}$, and elapsed time is measured from the trigger edge ($\mathrm{ET}=0$ on the edge scan), so a \code{TON} fires exactly $\lceil \mathrm{PT}/\Delta t\rceil$ scans after its input rises. From this one definition, K yields both the interpreter (\code{krun}) used below and the prover (\code{kprove}) used in \S\ref{sec:eval:rq4}.

\subsection{RQ1: An Executable Semantics Faithful to \gls{iec}~\mbox{61131-3}}
\label{sec:eval:rq1}

\subsubsection*{Metric} For each function block, whether \kesbmc{}'s per-scan done-bit sequence corresponds to the \openplc{}/\gls{matiec} reference on \emph{every} scan of the trace -- a per-block pass/fail, not an aggregate score.

\subsubsection*{Method: Checking Against \openplc{}} The behavioral reference for the function blocks is the C that \openplc{} actually executes: \gls{matiec}'s reference implementations (\code{iec\_std\_FB.h}) of \code{TON}, \code{TOF}, \code{TP}, \code{CTU}, and \code{CTD}. We extract those bodies unmodified into a small harness that, for a chosen input trace, zero-initializes the block instance and, on each scan, sets its inputs, advances the block's time input by a fixed $\Delta t$, invokes the reference body, and records the done bit. The traces exercise each block's characteristic behavior: an input held, then released (\code{TON}, \code{TOF}); a pulse (\code{TP}); and count/reset edge sequences (\code{CTU}, \code{CTD}). We then run \kesbmc{} on the same program, and the same trace, with the same $\Delta t$, and compare the two done-bit sequences scan for scan. Because the harness drives the block's own time input rather than wall-clock time, both engines see identical timing, and the comparison is exact and deterministic. Executing the \emph{same} reference code deployed on open-hardware \glspl{plc} means agreement confirms that \kesbmc{} matches real \gls{plc} behavior, not simply a paper specification.

\subsubsection*{Result: Exact Agreement} Table~\ref{tab:kesbmc:conformance} shows the outcome for the five function blocks (preset~$3$ for timers, $2$ for counters, $\Delta t=1$). \kesbmc{} reproduces the \openplc{}/\gls{matiec} done-bit trace \emph{exactly, scan for scan}, in every case. The combinational and latch constructs are likewise confirmed: an \code{AND} rung follows its contacts in both polarities, and a set/reset latch yields the retention trace $\langle\top,\top,\bot,\bot\rangle$ -- the light stays latched after its set input drops, exactly as \code{OTL}/\code{OTU} require.

\begin{table}[t]
    \centering
    \caption{RQ1 -- \kesbmc{} vs the \openplc{}/\gls{matiec} reference, done-bit trace per scan (timers $\mathrm{PT}=3$, counters $\mathrm{PV}=2$, $\Delta t=1$; $\top$/$\bot$ for true/false). \kesbmc{} equals the executed reference on every scan of every block.}
    \label{tab:kesbmc:conformance}
    \small
    \begin{tabular}{lcc}
        \toprule
        Block & Done-bit trace (\kesbmc{} $=$ \openplc{}) & Match \\
        \midrule
        \code{TON} (on-delay)  & $\bot\,\bot\,\bot\,\top\,\bot$                 & \cmark \\
        \code{TOF} (off-delay) & $\top\,\top\,\top\,\top\,\top\,\bot$           & \cmark \\
        \code{TP}  (pulse)     & $\bot\,\top\,\top\,\top\,\bot\,\top$           & \cmark \\
        \code{CTU} (count up)  & $\bot\,\bot\,\bot\,\bot\,\top\,\bot\,\bot$     & \cmark \\
        \code{CTD} (count down)& $\bot\,\bot\,\bot\,\bot\,\top\,\top\,\bot$     & \cmark \\
        \bottomrule
    \end{tabular}
\end{table}

\subsubsection*{The Validation is Discriminating} That the check is meaningful -- not a tautology -- is shown by the defects it caught while \kesbmc{} was being written. An early version of all three timers fired \emph{one scan too early}: it counted the trigger scan itself as one $\Delta t$ of elapsed time, whereas the standard (and \gls{matiec}) measure elapsed time from the trigger edge. A second version let \code{CTU} increment past its preset instead of saturating. Both differed from the reference and were flagged by this exact comparison; the timers were then rewritten as a direct port of the reference state machine, and \code{CTU} was capped, after which the traces coincide, as shown in Table~\ref{tab:kesbmc:conformance}. A validation that catches such off-by-one and saturation errors is a genuine test of fidelity and is what licenses the use of \kesbmc{} as the reference oracle in \S\ref{sec:e3}.

\subsection{RQ4: Machine-Checked Construct Lemmas}
\label{sec:eval:rq4}

\subsubsection*{Metric} The number of per-construct correctness lemmas \code{kprove} discharges against the semantics -- each either proved (\code{kprove} returns $\top$) or not -- and the fragment of \gls{ld} they cover. RQ1 grounds \kesbmc{} by \emph{execution}; RQ4 goes further and \emph{proves}, in \code{kprove}, that \kesbmc{}'s rules implement the \gls{iec}~\mbox{61131-3} input/output relation for each construct. We emphasize scope: we do not mechanize a full \kesbmc{}$\leftrightarrow$GOTO equivalence, which would require formalizing \gls{esbmc}'s \gls{ir} as well; the translation is validated empirically in \S\ref{sec:e3}, and RQ4 instead furnishes machine-checked correctness of the \emph{oracle itself}.

\subsubsection*{Per-Construct Correctness} For a construct, we state a reachability claim: from an image with symbolic bit values, executing the construct's rung yields exactly the standard-mandated coil value. Each is discharged against the semantics by \code{kprove} (Haskell backend). Table~\ref{tab:kesbmc:proof} lists the seven lemmas proved for the combinational and latch fragment -- contacts (series, parallel, negation), \code{OTE}, and \code{OTL}/\code{OTU} in both their acting and their \emph{retaining} cases -- all discharged (\code{kprove} returns $\top$).

\begin{lemma}[Per-Construct Correctness, Representative Case]
\label{lem:kesbmc:construct}
For all identifiers and all bit values $v_a,v_b$, executing the rung $\code{OTE}(q)\!:=\!\code{XIC}(a)\ \code{*}\ \code{XIC}(b)$ from an image binding $a\!\mapsto\!v_a,\ b\!\mapsto\!v_b$ terminates with $q\mapsto v_a\wedge v_b$; the analogous statements hold for parallel contacts ($\vee$), \code{XIO} ($\lnot$), and for \code{OTL}/\code{OTU} setting/resetting on a true rung and \emph{retaining} the coil on a false one.
\end{lemma}

\begin{table}[t]
    \centering
    \caption{RQ4 -- construct-correctness lemmas machine-checked in \code{kprove}. All seven discharge ($\top$).}
    \label{tab:kesbmc:proof}
    \small
    \begin{tabular}{@{}llc@{}}
        \toprule
        Lemma & Property proved & \code{kprove} \\
        \midrule
        L1 & series contacts $\Rightarrow$ coil $= \code{bit}(a)\wedge\code{bit}(b)$ & \cmark \\
        L2 & parallel contacts $\Rightarrow$ coil $= \code{bit}(a)\vee\code{bit}(b)$  & \cmark \\
        L3 & \code{XIO} $\Rightarrow$ coil $= \lnot\,\code{bit}(a)$                    & \cmark \\
        L4 & \code{OTL} sets the coil on a true rung                                  & \cmark \\
        L5 & \code{OTL} \emph{retains} the coil on a false rung                       & \cmark \\
        L6 & \code{OTU} resets the coil on a true rung                                & \cmark \\
        L7 & \code{OTU} \emph{retains} the coil on a false rung                       & \cmark \\
        \bottomrule
    \end{tabular}
\end{table}

\subsubsection*{Boundary of the Mechanized Guarantee} The timer and counter constructs are not yet mechanized: their correctness is a multi-scan property requiring induction over the input list, and the current scan-loop encoding is not discharged automatically by \code{kprove} (the prover does not narrow the symbolic input list into the constructor forms the scan rule matches, so the induction does not start). We state this plainly as the boundary of the proof: the combinational and latch fragment is \emph{machine-checked}, while the timer and counter constructs are \emph{execution-validated} against \openplc{} (RQ1) and \emph{empirically validated} through the differential study (\S\ref{sec:e3}). Closing the inductive lemmas -- by a list case-split lemma or a narrowing-friendly reformulation of the scan driver -- is the natural next step. It would lift the full fragment to an automated guarantee.

\subsubsection*{Threats to Validity (RQ1, RQ4)} For RQ1, the fidelity check exercises each block with representative traces rather than its full input space; we counter this by choosing traces that drive the discriminating behaviors (elapsed-time firing, saturation, edge counting), which is what surfaced the defects reported above. We take \gls{matiec}'s reference implementations as ground truth because they are the code actually executed on open-hardware \glspl{plc} -- the operative behavior a verifier must match -- and timing uses a controlled cycle time $\Delta t$, abstracting real-time jitter as the standard scan-cycle model does. For RQ4, the guarantee is scoped to the combinational and latch \emph{constructs} in isolation, not to whole programs, and the lemmas are stated for arbitrary distinct identifiers; their soundness rests on the trust base of the K reachability prover, as with any \code{kprove} result. These are the reasons RQ1 and RQ4 are complemented by -- not substituted for -- the end-to-end differential evidence of \S\ref{sec:e3}.

\section{Differential Validation of the LD\texorpdfstring{$\rightarrow$}{->}GOTO Translation}
\label{sec:e3}

Having established that \kesbmc{} is a faithful executable model of \gls{iec}~\mbox{61131-3} (\S\ref{sec:eval:rq1}), we use it as a \emph{reference oracle} for the \gls{esbmc} \gls{ld}$\rightarrow$GOTO translation. This section answers two questions: on the benchmark suite, does \gls{esbmc} agree with \kesbmc{} (RQ2), and where they disagree, which engine is correct and why (RQ3). Figure~\ref{fig:arch} shows the workflow.

\input{fig_arch}

\subsection{Experimental Setup}
\label{sec:e3:setup}

\subsubsection*{Benchmarks} We use the public \gls{esbmc}-\gls{plc} benchmark suite: 13 \gls{ld} programs drawn from eight controllers -- tank level control, bottle filling, elevator, traffic light, stairs light, water control, and two Beremiz examples --  several of which ship in a \textsc{safe} and a deliberately \textsc{unsafe} variant. Ten programs are combinational or latch controllers; the remaining ones exercise timers, counters, and edge blocks. The suite spans both diagram encodings: a \emph{simple} linear \code{<rung>} format and the full \emph{graphical} PLCopen netlist. Each program comes with a \code{props.yaml} file of safety properties specified with a \code{kind}: \code{invariant} (must always hold), \code{absence} (a bad condition that must never hold), or \code{mutual\_exclusion} (of a variable set, at most one true). 

One further example is a \gls{fbd} body and is out of scope; timer presets that a benchmark leaves unspecified are instantiated with a fixed default, which affects only \emph{when} a timer fires (\S\ref{sec:e3:threats}). Because the public suite exercises the construct fragment only incidentally -- its one timer is dead code (\S\ref{sec:e3:mutation}) -- we complement it with a controlled \emph{synthetic family} of 15 small programs (bringing the corpus to 28), each targeting one construct or interaction -- contact networks, latch and seal-in, \code{TON} chains, \code{TOF}/\code{TP} holds, \code{CTU}/\code{CTD} saturation, and edge detection -- with a property that \emph{observes} that construct. A coverage matrix (provided with the artifact) records that every construct and every property kind is exercised; this family drives the coverage and fault-injection study of \S\ref{sec:e3:mutation}.

\subsubsection*{Three Engines} Each program is checked by (i)~\gls{esbmc}, invoked as \code{esbmc <prog>.ld --ld-props <props>.yaml --unwind 6 --no-unwinding-assertions}, which returns \textsc{successful}/\textsc{failed} with a per-property counterexample on failure; (ii)~the \kesbmc{} interpreter (\code{krun}); and, for any disagreement, (iii)~\openplc{}'s runtime (\gls{matiec}-generated C), whose reference function-block semantics \kesbmc{} was validated against scan-for-scan in \S\ref{sec:eval:rq1}. Engine~(iii) is the external tie-breaker: it decides which of \gls{esbmc} and \kesbmc{} matches real \gls{plc} execution.

\subsubsection*{Two Independent Front-Ends} To keep \kesbmc{} an \emph{independent} oracle we never reuse \gls{esbmc}'s parser. A translator lowers each PLCopen~XML program into \kesbmc{} \gls{dsl}, reading the interface to classify variables as inputs, outputs, or locals (by section, or by the \code{\%IX}/\code{\%QX} located address). A \emph{simple} path handles the linear \code{<rung>} format (series contacts~$=$~AND, coils' \code{storage} set/reset~$\rightarrow$~latch/unlatch); a \emph{graphical} path resolves the netlist by tracing \code{refLocalId} wires backward from each coil -- power rails~$=$~\textsc{true}, chained contacts~$=$~AND, multiple incoming wires~$=$~OR -- and emits \code{TON}/\code{TOF}/\code{TP} and \code{CTU}/\code{CTD} blocks, with rising/falling edge contacts lowered to prev-value \code{R\_TRIG}/\code{F\_TRIG} helper rungs. The translators carry no verification semantics; they only re-express the diagram. We confirmed the graphical path against a manual translation of one program, obtaining an identical \kesbmc{} program.

\subsubsection*{The Differential Procedure} \gls{esbmc} verifies over a bounded scan horizon, so \kesbmc{} is exercised comparably and, crucially, its inputs are driven \emph{nondeterministically}. For a program with $k$ free inputs, we build one input \emph{trace}: the $2^{k}$ input combinations concatenated and repeated over three sweeps (so latch, timer, and counter state evolve across cycles); when $2^{k}>48$ we sample $48$ combinations, always including the all-false and all-true assignments. \kesbmc{} runs the translated program on this trace in a single execution, logging the complete variable image after every scan; we then evaluate each property on every logged image, interpreting the \code{kind} field exactly as \gls{esbmc}'s \code{--ld-props} does. A program is \textsc{unsafe} for \kesbmc{} iff some scan falsifies some property, and that scan is reported as the witness; otherwise it is \textsc{safe} up to the horizon. Running one long trace rather than per-transition queries amortizes interpreter start-up while still, at each scan, exploring the program's response to every input assignment -- a bounded exploration directly comparable to \gls{esbmc}'s \code{--unwind}. Timing uses one time unit per scan ($\Delta t = 1$).

\subsubsection*{Reproducibility} The \kesbmc{}, both front-ends, the differential driver, the \openplc{}/\gls{matiec} validation harness, and the \code{kprove} specifications are provided as an artifact; each row of Table~\ref{tab:e3} and each validation trace is reproduced by a single command, and the toolchain runs in a container so results do not depend on a local install. The driver also admits auxiliary corroboration engines -- the diagram compiled to C by \gls{matiec} and checked with \gls{cbmc}, and NuSMV on a \gls{bdd} encoding -- used in \S\ref{sec:e3:rq3} to double-check the timer mechanism with decision procedures independent of \kesbmc{}.

\subsection{RQ2: \kesbmc{} Agrees with \texorpdfstring{\gls{esbmc}}{ESBMC} on Well-Modeled Programs} 
\label{sec:e3:rq2}

\subsubsection*{Metric} The fraction of programs on which \kesbmc{} and \gls{esbmc} return the same \textsc{safe}/\textsc{unsafe} verdict, and -- on the \textsc{unsafe} programs --  whether \kesbmc{} additionally localizes the violation to specific properties with a witness.

Table~\ref{tab:e3} reports all 13 benchmark programs. On \textbf{10 of 13}, the two engines return the same verdict. The agreements span the full construct fragment: combinational interlocks (\code{tank\_level\_control}), sealed-in logic (\code{water\_control}), and timer- and edge-driven controllers (\code{bottle\_filling}, \code{elevator}, \code{beremiz\_traffic\_light}). Crucially, on every \textsc{unsafe} variant \kesbmc{} not only matches \gls{esbmc}'s pass/fail but \emph{localises} the violation to specific properties with a concrete witness image -- for example, the unsafe \code{bottle\_filling} flags \code{P1}\,(valve open with no bottle) and \code{P2}\,(valve open when full), while correctly keeping \code{P3},\code{P5} safe because the emergency-stop guard still propagates through \code{System\_Running} within a scan. This per-property, witness-bearing verdict is strictly more informative than \gls{esbmc}'s aggregate result.

\begin{table}[t]
    \centering
    \caption{Three-engine differential over the benchmark suite. \kesbmc{} verdicts list the violated properties. Rows marked \textbf{\xmark} disagree; all three are resolved in \kesbmc{}'s favor in \S\ref{sec:e3:rq3}.}
    \label{tab:e3}
        \small
        \begin{tabular}{@{}llll c@{}}
        \toprule
        Program & Fmt & \kesbmc{} & \gls{esbmc} & Agree \\
        \midrule
        tank\_level\_control          & rung & \textsc{safe}                 & \textsc{safe}   & \cmark \\
        tank\_level\_control\,(unsafe)& rung & \textsc{unsafe} P1,P2,P3      & \textsc{unsafe} & \cmark \\
        bottle\_filling               & rung & \textsc{safe}                 & \textsc{safe}   & \cmark \\
        bottle\_filling\,(unsafe)     & rung & \textsc{unsafe} P1,P2,P4,P6   & \textsc{unsafe} & \cmark \\
        elevator                      & rung & \textsc{safe}                 & \textsc{safe}   & \cmark \\
        elevator\,(unsafe)            & rung & \textsc{unsafe} P1--P6        & \textsc{unsafe} & \cmark \\
        traffic\_light\,(safe)        & rung & \textsc{safe}                 & \textsc{safe}   & \cmark \\
        water\_control                & graph & \textsc{safe}                & \textsc{safe}   & \cmark \\
        beremiz\_traffic\_light       & graph & \textsc{safe}                & \textsc{safe}   & \cmark \\
        dimmer\_light\_control        & graph & \textsc{safe}                & \textsc{safe}   & \cmark \\
        \midrule
        \textbf{stairs\_light}        & graph & \textbf{\textsc{unsafe} P1}   & \textsc{safe}   & \xmark \\
        \textbf{traffic\_light}       & rung & \textsc{safe}                 & \textbf{\textsc{unsafe}} & \xmark \\
        \textbf{traffic\_light\,(unsafe)} & rung & \textsc{safe}             & \textbf{\textsc{unsafe}} & \xmark \\
        \bottomrule
    \end{tabular}
\end{table}

\subsection{RQ3: Every Disagreement Is an \gls{esbmc} Timer Defect}
\label{sec:e3:rq3}

\subsubsection*{Metric} For each disagreement, which engine-independent execution (\openplc{}) confirms, and its root cause -- yielding a classification of the translation defects the oracle exposes. The three disagreeing runs -- \code{traffic\_light} in both its safe and unsafe variants, plus the graphical \code{stairs\_light} -- reduce to two defect classes on two distinct programs, and all involve timer function blocks; \openplc{} confirms \kesbmc{} in every case, and on the controlled probe below two further verification engines corroborate \kesbmc{} as well (Table~\ref{tab:e3:multiengine}). They expose two \emph{opposite} defects in \gls{esbmc}'s incomplete timer translation (Table~\ref{tab:e3tax}).

\subsubsection*{A Controlled Probe} We first isolate the mechanism with a minimal program: \code{Btn}$\rightarrow$\code{TON (PT)}$\rightarrow$\code{Light}. Under a faithful timer, the invariant \code{A}: ``\code{!Light || Btn}'' holds -- \code{Light} follows the done bit, and a \code{TON} holds its output only while its input is high. \kesbmc{} proves \code{A} safe; \gls{esbmc} reports it \emph{violated}, with a counterexample in which the done bit is true while the input is false -- a state a \code{TON} can never reach, since a low input resets it. \gls{esbmc} therefore leaves the timer output \emph{unconstrained} (havoc), an over-approximation that manufactures false alarms.

\subsubsection*{Independent Cross-Engine Corroboration} The probe's verdict does not rest on \kesbmc{} alone. We re-check the invariant \code{A} with two engines that share no code with either \gls{esbmc} or \kesbmc{} and that use different decision procedures: the diagram compiled to C by \gls{matiec} and discharged by \gls{cbmc}~\cite{Clarke2004cbmc} -- a distinct SAT-based \gls{bmc} backend -- and NuSMV~\cite{Cimatti2002nusmv}, whose \gls{bdd}-based symbolic model checking is unrelated to bounded unrolling. Both report \code{A} \textsc{safe}, and NuSMV's proof is \emph{unbounded} -- a fixpoint over all reachable states, not simply up to a horizon. Together with the \openplc{} runtime, four accounts -- differing in decision procedure -- thus agree against \gls{esbmc} (Table~\ref{tab:e3:multiengine}), isolating its timer \emph{havoc} as the sole cause of the false alarm. 

We are precise about what this independence remains and is not: it rules out an artifact of \kesbmc{}'s modeling or of \gls{bmc} in particular, but it is not a fully independent second account of the standard -- the \openplc{} and \gls{matiec}--\gls{cbmc} engines both execute \gls{matiec}'s function-block semantics (against which \kesbmc{} was itself validated, \S\ref{sec:eval:rq1}), and the NuSMV model encodes the same faithful timer. \gls{esbmc} is the outlier, and the only engine here with an independent source-to-model front-end. Beyond the probe, NuSMV independently confirms the two decisive benchmarks themselves: on faithful models it proves \code{traffic\_light} \textsc{safe} (the green-conflict mutual exclusion holds, by an unbounded fixpoint) and finds \code{stairs\_light} \textsc{unsafe} (\code{P1} violated), matching \kesbmc{} against \gls{esbmc} on the real programs -- not only the abstract probe. The differential harness also admits these engines to the remaining benchmarks.

\begin{table}[t]
    \centering
    \caption{Controlled timer probe \code{Btn}$\to$\code{TON}$\to$\code{Light}, invariant \code{A}: ``\code{!Light || Btn}''. Four independent accounts -- an executable oracle, two verifiers with unrelated decision procedures, and the deployed runtime -- agree the property holds under a faithful timer; only \gls{esbmc}, which \emph{havocs} the timer output, reports a violation. NuSMV's \gls{bdd} proof is unbounded.}
    \label{tab:e3:multiengine}
    \small
    \begin{tabular}{@{}llc@{}}
        \toprule
        Engine & Method & \code{A} \\
        \midrule
        \gls{esbmc}-\gls{plc}          & SMT-\gls{bmc} (havocs timer)   & \textsc{violated} \\
        \kesbmc{}                       & executable reachability        & \textsc{safe} \\
        \gls{matiec}$\to$C\,$+$\,\gls{cbmc} & SAT-\gls{bmc}              & \textsc{safe} \\
        NuSMV                           & \gls{bdd} symbolic (unbounded)       & \textsc{safe} \\
        \openplc{}/\gls{matiec}            & concrete execution             & \textsc{safe} \\
        \bottomrule
    \end{tabular}
\end{table}

\paragraph{Discrepancy 1 -- Havoc False Alarms (\code{traffic\_light}).} The real \code{traffic\_light} controller sequences four phases through a chain of four \code{TON}s. \gls{esbmc} declares it \textsc{unsafe} on the green-conflict mutual-exclusion property; its counterexample asserts \emph{all four phases active in the same scan}. A real \code{TON} chain fires one phase at a time, so this state is unreachable -- it arises only because \gls{esbmc} havocs the four done bits. Executed under \kesbmc{}, the controller exhibits \emph{zero} mutual-exclusion violations across a full multi-cycle horizon (indeed it deadlocks in the north--south phase, a liveness defect that violates no \emph{safety} property). \kesbmc{}'s \textsc{safe} verdict is correct, and independent of the timer preset.

\paragraph{Discrepancy 2 -- Skipped Rimers Miss Real Bugs (\code{stairs\_light}).} The graphical \code{stairs\_light} benchmark drives its light through an off-delay (\code{TOF}) so that a single PIR detection keeps the light on for the preset. Its front-end warns that ``FB/timer outputs on rail$\rightarrow$coil paths are not yet modeled'' and \emph{skips} the timer path, after which \gls{esbmc} certifies the program \textsc{safe}. But that skip is exactly where the violation lives: once the PIR clears while the light is still held on by the off-delay, invariant \code{P1} (``light on implies PIR active or button state on'') is falsified. \kesbmc{}, modeling the \code{TOF}, reports \code{P1} \textsc{unsafe} with the light-on/PIR-off witness, and \openplc{} reproduces the same light activity. Here \gls{esbmc} is \emph{unsound}: it returns \textsc{safe} for a program that violates its own stated property.

\begin{table}[t]
    \centering
    \caption{Discrepancy taxonomy. Every disagreement is an \gls{esbmc} timer-translation defect; \kesbmc{} is correct in all three, corroborated by \openplc{}.}
    \label{tab:e3tax}
    \small
    \begin{tabular}{@{}lllp{8cm}@{}}
        \toprule
        Program & \gls{esbmc} & \kesbmc{} & \gls{esbmc} defect \\
        \midrule
        \code{stairs\_light}   & \textsc{safe}   & \textsc{unsafe} & \emph{skips} the FB path (graphical) $\rightarrow$ under-approx.\ $\rightarrow$ \textbf{missed bug} (unsound) \\
        \code{traffic\_light}  & \textsc{unsafe} & \textsc{safe}   & \emph{havocs} the FB output (rung) $\rightarrow$ over-approx.\ $\rightarrow$ \textbf{false alarm} \\
        \bottomrule
    \end{tabular}
\end{table}

\subsection{Fault Injection: The Property-Adequacy Gap}
\label{sec:e3:mutation}

A clean agreement (RQ2) is only as strong as the properties that could have exposed a disagreement. To examine this, we inject single-point \emph{translation-rule} faults into the benchmark programs and ask two questions of each mutant: does it change the program's observable behavior -- the per-scan trace \kesbmc{} computes -- and does it flip any of the benchmark's own safety properties? The gap between the two answers is the finding.

\subsubsection*{Method} We define five mutation operators, each emulating a class of front-end fault: contact-polarity swap (\code{XIC}$\leftrightarrow$\code{XIO}), coil-operation change (\code{OTE}$\rightarrow$\code{OTL}), guard-contact drop (remove one series contact), latch-retention drop (\code{OTL}/\code{OTU}$\rightarrow$\code{OTE}), and timer-kind swap (\code{TON}$\leftrightarrow$\code{TOF}). We generate every single-point mutant of a combinational benchmark (\code{tank\_level\_control}) and a timer benchmark (\code{bottle\_filling}) -- 60 mutants -- and classify each as \emph{behaviorally detected} if its trace differs from the baseline, and as \emph{property-detected} if it flips a benchmark safety property.

\subsubsection*{Result} \kesbmc{} distinguishes \emph{every} behaviorally distinct mutant: 40 of the 60 mutants (67\%) change the observable trace, and the oracle flags all of them; the remaining 20 are semantically equivalent (no observable effect, the ceiling for any behavioral method). The benchmark \emph{properties}, however, detect only \textbf{11 of those 40 behavior-changing faults -- 27\%}. In other words, most genuine translation faults alter behavior that the shipped safety properties never observe: they touch local variables the properties do not constrain, restrict an output in a direction a ``must-not'' property cannot see, or -- strikingly -- affect \emph{dead code}. In \code{bottle\_filling}, the timer's output \code{Filling\_Done} is overwritten by a later rung, so the \code{TON}$\rightarrow$\code{TOF} swap is behaviorally equivalent and no property moves; this is also precisely why \gls{esbmc}'s skipping that timer is harmless \emph{for this program}.

\subsubsection*{Why this Matters} The experiment measures \emph{property adequacy}, not oracle sensitivity: the oracle sees every observable fault, but a property-based verifier catches a translation defect only when a property happens to constrain the affected variable. This is exactly the blind spot that lets defects slip through -- and it explains the pattern in \S\ref{sec:e3:rq3}. When the benchmark properties observe the timer (\code{stairs\_light}, \code{traffic\_light}), the oracle exposes the \gls{esbmc} defects; when they do not (\code{tank}, \code{bottle}), a timer-translation fault passes silently. An executable oracle that exposes the \emph{whole} behavior, rather than only property verdicts, is what makes such faults visible at all. (Two benchmarks are used here; the latch operator does not fire, as neither uses latched coils, and the timer operator fires once on a dead timer -- so the timer/latch classes are under-exercised, which the synthetic family below addresses.)

\subsubsection*{A Systematic Campaign on the Synthetic Family} The two-benchmark study under-exercises the latch and timer operators; the controlled family (\S\ref{sec:e3:setup}) closes that gap. Running the same five operators over its latch- and timer-heavy programs yields \textbf{87 mutants} in which the previously inert operators now fire -- latch-retention \textbf{four} times (was \emph{zero}) and timer-kind \textbf{five} times (was one, on dead code). \kesbmc{} again distinguishes every behaviorally distinct mutant ($73$ of $87$ change the trace); the shipped properties detect \textbf{$32$ of those $73$ -- $44\%$} (Table~\ref{tab:e3:family}). The gap persists even here, where all property was \emph{designed} to observe its construct: property-detection was $27\%$ on the public benchmarks and $44\%$ on the purpose-built family -- in both cases, well under half of the genuine behavioral faults. Property adequacy, not oracle sensitivity, remains the limiting factor.

\begin{table}[t]
    \centering
    \caption{Fault injection over the synthetic family (\S\ref{sec:e3:setup}): the five translation-fault operators, now all firing. \kesbmc{} detects every behaviorally distinct mutant; the shipped properties detect $44\%$ of them.}
    \label{tab:e3:family}
    \small
    \begin{tabular}{@{}llccc@{}}
        \toprule
        Op & Fault class & Mutants & Behavioral & Property \\
        \midrule
        CP & contact polarity  & 49 & 39 & 17 \\
        CO & coil operation    & 20 & 20 & 8  \\
        LR & latch retention   & 4  & 4  & 2  \\
        GC & guard drop        & 9  & 6  & 3  \\
        TK & timer kind        & 5  & 4  & 2  \\
        \midrule
        \textbf{Total} & & \textbf{87} & \textbf{73} & \textbf{32 ($44\%$)} \\
        \bottomrule
    \end{tabular}
\end{table}

\subsection{Discussion and Threats}
\label{sec:e3:threats}

\subsubsection*{What the Oracle Bought} The differential is decisive precisely because \kesbmc{} is an \emph{executable} \gls{iec}-faithful semantics: it does not simply flag a mismatch, it produces the concrete scan-level behavior that shows which engine is right, and \openplc{} confirms it. The two defects are complementary, and both are practically serious -- the havoc mode wastes engineering effort on impossible counterexamples. In contrast, the skip mode is unsound and can certify a hazardous program safe.

\subsubsection*{Threats to Validity} (i)~The comparison is bounded and input-sampled. Hence, a \textsc{safe}/\textsc{safe} \emph{agreement} is agreement \emph{up to the horizon}, not a proof of joint correctness: in principle, both engines could miss the same violation inside the bound. We reduce this in two ways. Our conclusions rest on the \emph{disagreements}, not the agreements, and each disagreement is settled by \openplc{} and a preset-independent structural argument rather than by the bounded run alone. The input trace exercises every input assignment at every scan over several cycles, so a missed violation would have to evade the entire enumerated input space at every logged scan. Reported \textsc{safe} verdicts should nonetheless be read as safe-up-to-horizon, matching \gls{esbmc}'s own bounded model -- with one exception at the heart of the disagreements: on the controlled timer probe (\S\ref{sec:e3:rq3}) NuSMV discharges invariant \code{A} by an \emph{unbounded} \gls{bdd} fixpoint, so there the \textsc{safe} verdict holds over all reachable states, not only to the horizon. (ii)~Timer presets absent from a benchmark are instantiated with a fixed default; this affects \emph{when} a timer fires, not the existence of the discrepancies, each of which we tie to a structural argument. (iii)~The two front-ends are independent of \gls{esbmc} and were cross-checked by reproducing a manually translated benchmark identically, so a divergence cannot be an artifact of a shared parser. (iv)~\emph{External validity.} The study covers 13 programs and one verifier; the specific defects may not generalize, but the oracle and harness are reusable on any \gls{ld} program and any verifier (\S\ref{sec:limits}).

\section{Case Studies}
\label{sec:cases}

To make the framework concrete, we walk through the two disagreements of \S\ref{sec:e3:rq3} end to end: from the benchmark diagram, through its independent translation into \kesbmc{}, to the per-scan execution that settles which verifier is right. Each illustrates one of \gls{esbmc}'s timer-translation failure modes.

\subsection{Missed Violation: the Stairs Light (unsound \emph{skip})}
\label{sec:cases:stairs}

The \code{stairs\_light} controller (graphical format) turns a stairwell light on when a passer-by is detected and, via an off-delay timer, \emph{keeps} it on for a preset afterward. Listing~\ref{lst:stairs} is the program our graphical front-end produces from the PLCopen netlist, with no reference to \gls{esbmc}. Its last two logic rungs are the crux: the off-delay \code{TOF0\_Q} is driven by a rising edge of the PIR sensor, and the light is \code{TOF0\_Q} \emph{or} the manual button state.

\begin{lstlisting}[caption={\kesbmc{} program for \code{stairs\_light}, produced by the graphical front-end (edge and previous-value helper rungs elided).}, label={lst:stairs}]
TOF(TOF0_Q, 20, XIC(rise_stairs_pir_sensor) * XIO(lights_buttons_state)) ;
OTE(stairs_light) := XIC(TOF0_Q) + XIC(lights_buttons_state) ;
\end{lstlisting}

The benchmark's property \code{P1} states that the light is on only if the PIR is active or the button state is set: \code{!stairs\_light || stairs\_pir\_sensor || lights\_buttons\_state}. We execute \kesbmc{} on a single PIR pulse alongside the buttons untouched; Table~\ref{tab:stairs} shows the resulting per-scan image.

\begin{table}[t]
    \centering
    \caption{\code{stairs\_light} under \kesbmc{}: a PIR pulse and no button pressed. From scan~2, the off-delay holds the light on while the corridor is empty, falsifying \code{P1}.}
    \label{tab:stairs}
    \small
    \begin{tabular}{ccccc}
        \toprule
        Scan & \code{PIR} & \code{buttons} & \code{light} & \code{P1} \\
        \midrule
        0 & $\bot$ & $\bot$ & $\bot$ & holds \\
        1 & $\top$ & $\bot$ & $\top$ & holds (PIR active) \\
        2 & $\bot$ & $\bot$ & $\top$ & \textbf{violated} \\
        3 & $\bot$ & $\bot$ & $\top$ & \textbf{violated} \\
        4 & $\bot$ & $\bot$ & $\top$ & \textbf{violated} \\
        \bottomrule
    \end{tabular}
\end{table}

At scan~1, the PIR fires and the light comes on -- legitimately, the sensor is active. At scan~2 the person has left (\code{PIR}~$=\bot$) and no button is latched (\code{buttons}~$=\bot$), yet the off-delay still holds \code{light}~$=\top$: this falsifies \code{P1}, and \kesbmc{} reports the violation with scan~2 as the witness. \gls{esbmc}, however, returns \textsc{verification successful} on this exact program: its front-end warns that the timer block is ``not yet modeled'' and \emph{skips} that rung, after which it never sees the light turn on at all. Because \kesbmc{}'s \code{TOF} was validated scan-for-scan against \openplc{} (\S\ref{sec:eval:rq1}), and \openplc{} reproduces the same similar behavior, the violation is real: \gls{esbmc} has certified as safe a program that violates its own stated property -- an \emph{unsound} outcome, the most dangerous kind for a safety tool.

\subsection{False Alarm: the Traffic Light (imprecise \emph{havoc})}
\label{sec:cases:traffic}

The \code{traffic\_light} controller sequences four phases -- north--south green and yellow, then east--west green and yellow -- through a chain of four on-delay timers, each phase enabling the next (Listing~\ref{lst:traffic}). A mutual-exclusion property requires that the two green phases never coincide: \code{!(NS\_Green \&\& EW\_Green)}.

\begin{lstlisting}[caption={\kesbmc{} program for \code{traffic\_light}: a chain of four \code{TON}s sequences the phases (excerpt).}, label={lst:traffic}]
TON(TON1_Q, 2, XIC(TON1_IN)) ;   OTE(Phase_NS_Green)  := XIC(TON1_IN) ;
TON(TON2_Q, 2, XIC(TON2_IN)) ;   OTE(Phase_NS_Yellow) := XIC(TON2_IN) ;
TON(TON3_Q, 2, XIC(TON3_IN)) ;   OTE(Phase_EW_Green)  := XIC(TON3_IN) ;
TON(TON4_Q, 2, XIC(TON4_IN)) ;   OTE(Phase_EW_Yellow) := XIC(TON4_IN) ;
OTE(NS_Green) := XIC(Phase_NS_Green) * XIO(Emergency_Vehicle) ;
OTE(EW_Green) := XIC(Phase_EW_Green) ;
\end{lstlisting}

\gls{esbmc} declares this program \textsc{unsafe} on the green-conflict property. Its counterexample (Table~\ref{tab:traffic}, left) asserts that all four phase flags are set in the same scan, so that \code{NS\_Green} and \code{EW\_Green} hold together. But that state is unreachable for a real timer chain: the phases are enabled in sequence, one at a time. \gls{esbmc} reaches it only because, on the simple diagram format, it leaves the four done bits \code{TON1\_Q}--\code{TON4\_Q} \emph{unconstrained} (havoc). Executing the program under \kesbmc{} with \code{Enable} held (Table~\ref{tab:traffic}, right) shows the true behavior: exactly one phase is active at a time and \code{NS\_Green}, \code{EW\_Green} are never simultaneously set -- no mutual-exclusion violation occurs over the whole cycle. \kesbmc{}'s \textsc{safe} verdict is correct, and independent of the timer preset. \gls{esbmc}'s counterexample is a modeling artifact -- a false alarm that no real timer can produce.

\begin{table}[t]
    \centering
    \caption{\code{traffic\_light}, green phases: \gls{esbmc}'s counterexample versus the actual \kesbmc{} execution. \gls{esbmc} needs all phases active at once (from havoc'd timer outputs); \kesbmc{} keeps the phases one-hot.}
    \label{tab:traffic}
    \small
    \begin{tabular}{@{}c cc c cc@{}}
        \toprule
        & \multicolumn{2}{c}{\gls{esbmc} (havoc CEX)} & & \multicolumn{2}{c}{\kesbmc{} (executed)} \\
        \cmidrule(lr){2-3}\cmidrule(lr){5-6}
        Scan & \code{NS\_Green} & \code{EW\_Green} & & \code{NS\_Green} & \code{EW\_Green} \\
        \midrule
        $s$   & $\top$ & $\top$ & & $\top$ & $\bot$ \\
        $s{+}1$ & --   & --     & & $\bot$ & $\bot$ \\
        $s{+}2$ & --   & --     & & $\bot$ & $\bot$ \\
        \midrule
        conflict & \multicolumn{2}{c}{\textbf{claimed}} & & \multicolumn{2}{c}{\textbf{never}} \\
        \bottomrule
    \end{tabular}
\end{table}

\subsection{What the Two Cases Show}
\label{sec:cases:summary}

The two cases are mirror images. In the stairs light \gls{esbmc} models \emph{too little} -- it drops the timer and misses a real violation (unsound). At the traffic light, it admits \emph{too much} -- it interrupts the timer and invents a violation that cannot occur (imprecise). Neither is visible from inside \gls{esbmc}: a \textsc{safe} verdict and an \textsc{unsafe} counterexample look the same whether or not the timer was modeled faithfully. What separates the true result from the artifact in each case is the same thing -- an executable, standard-faithful semantics that can be \emph{run} to produce the actual scan-by-scan behavior, and cross-checked against a real \gls{plc} runtime. That is the role \kesbmc{} plays, and these two controllers show it playing it.

\section{Discussion}
\label{sec:discussion}

\subsubsection*{What an executable reference bought} The central lesson of \S\ref{sec:e3} is not the raw count of discrepancies yet \emph{how} they were settled. A disagreement between two verifiers, or between a verifier and a paper specification, is inconclusive: either side might be wrong, and neither can be run. Because \kesbmc{} is \emph{executable} and validated against the very C that \openplc{} runs, each disagreement resolves to a concrete scan-level trace that shows which engine departs from \gls{iec}~\mbox{61131-3} -- and \openplc{} then reproduces that trace as an independent third witness. An executable, standard-faithful semantics is thus not simply another checker but an \emph{arbiter}: it converts ``the tools disagree'' into ``here is the scan on which the translation violates the standard.''

\subsubsection*{Two failure modes, both practically serious} \gls{esbmc}'s incomplete timer translation fails in opposite directions, and the distinction matters to an engineer. In graphical diagrams, it \emph{skips} the timer path and can therefore certify as safe a program that, in fact, violates its own stated property -- an \emph{unsound} outcome, the most dangerous kind for a safety tool, because the user receives a green light for a hazardous controller. On the simple format, it instead \emph{havocs} the timer output and manufactures counterexamples that no real timer can produce -- a false alarm that, while sound, erodes trust and wastes triage effort on impossible scenarios. Neither is visible from inside the tool: only an external account of what the diagram means can separate a true defect from a modeling artifact, which is precisely the role \kesbmc{} plays.

\subsubsection*{A layered correctness argument} \kesbmc{} supports a graded notion of assurance that we are careful to keep honest. The combinational and latch fragment is \emph{machine-checked} in \code{kprove} (\S\ref{sec:eval:rq4}); the timer and counter function blocks are \emph{execution-validated} scan-for-scan against the \openplc{}/\gls{matiec} reference (\S\ref{sec:eval:rq1}); and the end-to-end \gls{ld}$\rightarrow$GOTO translation is \emph{empirically validated} by the differential study (\S\ref{sec:e3}). Each layer is weaker than the one before but covers more, and together they close the \gls{esbmc}-\gls{plc} trust gap far more tightly than the prior state, in which the translation's fidelity rested only on its having been \emph{designed to match} an informal formalization.

\subsubsection*{Generality} Nothing in the method is specific to \gls{esbmc}. \kesbmc{} is a stand-alone reference interpreter for \gls{iec}~\mbox{61131-3} \gls{ld}; the differential harness compares \emph{any} verifier's verdicts against \kesbmc{}'s ground truth, and the two independent front-ends read the same PLCopen~XML that the tools consume. The same instrument could scrutinize other \gls{ld} verifiers, regression-test a verifier's front-end across releases, or serve as the executable oracle in a conformance test suite for \gls{iec}~\mbox{61131-3} tooling.

\subsubsection*{Impact} A formal verifier is itself an unverified program, and its soundness rests on a translation that has, until now, had no independent executable account of the source language to be checked against. The transferable contribution of this work is a way to close that loop: build an executable, standard-faithful reference semantics, use it as a differential oracle to \emph{audit} a translation against real execution, and \emph{machine-check} the reference where tractable -- so that trust in a toolchain becomes something one tests, adjudicates, and in part proves rather than assumes. The pattern applies wherever a tool verifies a language by lowering it into a model checker, a setting common far beyond \glspl{plc}. 

The demonstration also leaves reusable assets for distinct communities: an executable, \openplc{}-validated semantics of \gls{iec}~\mbox{61131-3} \gls{ld} for control-systems and cyber-physical-systems research, which has largely lacked one; a case study for the semantics-engineering community of a single K definition serving \emph{both} as the interpreter that tests another tool and the prover that certifies constructs; and, for the testing and certification communities, a concrete instance of finding a genuine \emph{soundness} defect -- a verifier reporting safe a program that is not -- by differential testing against a formal reference. In a field that increasingly relies on automated verifiers for safety cases, an oracle that can hold those verifiers to the standard constitutes a step toward auditable, rather than merely asserted, assurance.

\section{Limitations and Future Work}
\label{sec:limits}

\subsubsection*{Mechanization boundary} Our machine-checked lemmas cover the combinational and latch fragment. The timer and counter constructs are multi-scan properties whose proofs require induction over the input sequence, and the current scan-loop encoding is not discharged automatically by \code{kprove} (\S\ref{sec:eval:rq4}). Closing these lemmas -- via an explicit list case-split lemma or a narrowing-friendly reformulation of the scan driver -- would raise the whole section from execution-validated to machine-checked, and is the most direct extension of this work. A full, mechanized \kesbmc{}$\leftrightarrow$GOTO equivalence would additionally require a formal semantics of \gls{esbmc}'s \gls{ir}; our differential approach deliberately trades that heavyweight goal for high, reproducible coverage.

\subsubsection*{Bounded differential and presets} Like the \gls{bmc} it validates, the differential is bounded: \textsc{safe} verdicts are safe-up-to-horizon. Where a benchmark left a timer preset unspecified, we instantiated a fixed default, which affects \emph{when} a timer fires but not the existence of the reported discrepancies -- each of which we tie to a preset-independent structural argument. Extending to boundless guarantees (e.g.\ via $k$-induction over \kesbmc{} itself) is future work.

\subsubsection*{Coverage and external validity} Our conclusions rest on 13 differential benchmarks -- complemented by a 15-program synthetic family (corpus~28) for construct coverage and fault injection (\S\ref{sec:e3:mutation}) -- so the specific defects found may not generalize; the mitigating factor is that the oracle is \emph{reusable}, and we invite that reuse. Any \gls{iec}~\mbox{61131-3} \gls{ld} program in the covered fragment -- Boolean signals with contacts, coils and latches, \code{TON}/\code{TOF}/\code{TP}, \code{CTU}/\code{CTD}, and \code{R\_TRIG}/\code{F\_TRIG}, in either diagram format -- can be adjudicated against \kesbmc{} by translating it and running the differential (the artifact does this in one command per program), and against \emph{any} verifier, not only \gls{esbmc}. Outside the fragment -- non-Boolean data such as integer or analog arithmetic, non-standard or vendor function blocks (one such block appears in a single benchmark), and Function Block Diagram or Sequential Function Chart bodies (one benchmark) -- the front-end flags the unsupported construct instead of returning a wrong answer, and support is added rule by rule in the K definition. Broadening \kesbmc{} toward richer data types, and integrating it with the \kst{} Structured Text semantics for mixed-language projects, would widen the class of programs the oracle can adjudicate; until then, the covered fragment is precisely the limit within which a practitioner can apply the tool to their own programs today.

\subsubsection*{Fault injection} Beyond the two-benchmark study (\S\ref{sec:e3:mutation}), we ran the same operators over a controlled synthetic family (15 programs, corpus~28) that exercises every construct, observing properties, so that all five operators fire -- including the latch and timer operators the public benchmarks left inert (Table~\ref{tab:e3:family}). The property-adequacy gap holds there too: only $44\%$ of behavior-changing mutants flip a property, despite properties purpose-built to observe their construct. The remaining extension is to mutate a verifier's \emph{translation code} directly rather than the diagram, which requires access to each verifier's front-end source code. That the safety properties shipped with public benchmarks observe so little of a program's behavior is itself a call for stronger property suites, independent of any single verifier.

\section{Conclusion}
\label{sec:conclusion}

Automated verifiers for \gls{iec}~\mbox{61131-3} Ladder Diagram are only as trustworthy as the unverified front-end that lowers a diagram into a model checker. Yet that translation has had no independent, executable account of the standard against which to check it. We presented \kesbmc{}, an executable formal semantics of \gls{iec}~\mbox{61131-3} \gls{ld} in the K framework, extending \kst{} to the graphical language and its function blocks, from which a single definition yields both an interpreter and a prover. We validated \kesbmc{} scan-for-scan against the \openplc{}/\gls{matiec} reference implementations, used it as an independent oracle to validate the \gls{esbmc} differentially-\gls{plc} \gls{ld}$\rightarrow$GOTO translation -- finding broad agreement, witnessed violations on every unsafe program, and three corroborated defects that fall into two opposite timer-handling failure modes -- and machine-checked in \code{kprove} that \kesbmc{}'s rules implement the standard's input/output relation for the combinational and latch fragment. The wider contribution is methodological: a verifier's front-end is trusted code, and an executable, standard-faithful reference semantics turns that trust into something one can test, adjudicate, and, in part, prove. \kesbmc{} is a reusable instrument for doing so, and we hope it functions as an executable reference for \gls{iec}~\mbox{61131-3} \gls{ld} beyond the single verifier examined here.

%% file: fig_kesbmc_overview.tex
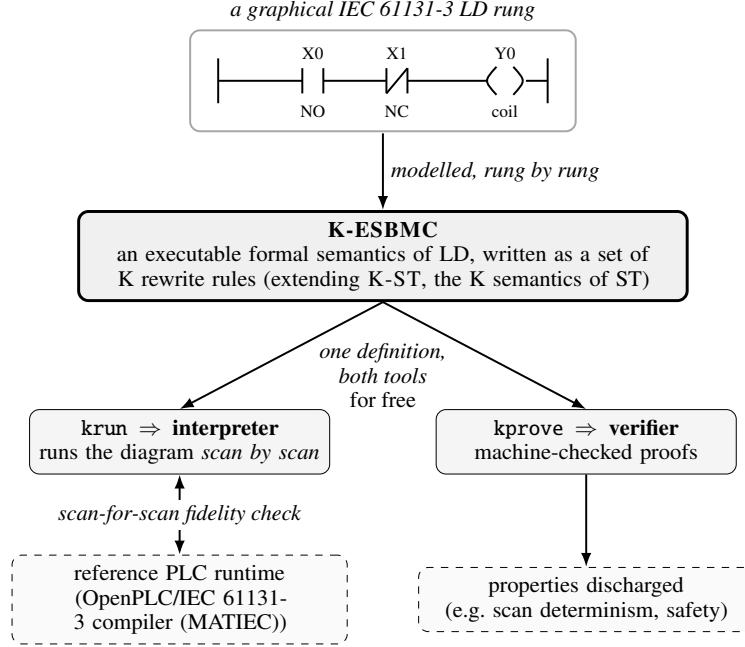
\begin{figure}[htbp]
    \centering
    \resizebox{0.6\linewidth}{!}{%
    \begin{tikzpicture}[
      font=\small,
      >={Latex[length=2mm]},
      box/.style   = {draw, rounded corners, align=center, inner sep=4pt, minimum height=8mm},
      defn/.style  = {box, very thick, fill=black!6},
      tool/.style  = {box, fill=black!4},
      ext/.style   = {box, dashed, fill=black!2},
      tag/.style   = {font=\footnotesize\itshape, inner sep=2pt},
      flow/.style  = {->, thick},
    ]

    \begin{scope}[shift={(0,4)}, line width=0.8pt]
      \draw (-2.35,-0.32) -- (-2.35,0.32);
      \draw ( 2.35,-0.32) -- ( 2.35,0.32);
      \draw (-2.35,0) -- (-1.15,0);
      \draw (-1.15,-0.18) -- (-1.15,0.18);
      \draw (-0.85,-0.18) -- (-0.85,0.18);
      \node[font=\scriptsize] at (-1.0,0.40) {X0};
      \node[font=\scriptsize] at (-1.0,-0.44) {NO};
      \draw (-0.85,0) -- (0.05,0);
      \draw (0.05,-0.18) -- (0.05,0.18);
      \draw (0.35,-0.18) -- (0.35,0.18);
      \draw (0.05,-0.18) -- (0.35,0.18);            
      \node[font=\scriptsize] at (0.20,0.40) {X1};
      \node[font=\scriptsize] at (0.20,-0.44) {NC};
      \draw (0.35,0) -- (1.50,0);
      \draw (1.62,0.20) .. controls (1.44,0) .. (1.62,-0.20);   
      \draw (1.86,0.20) .. controls (2.04,0) .. (1.86,-0.20);   
      \draw (2.04,0) -- (2.35,0);
      \node[font=\scriptsize] at (1.74,0.40) {Y0};
      \node[font=\scriptsize] at (1.74,-0.44) {coil};
      \draw[rounded corners, black!35] (-2.75,-0.72) rectangle (2.75,0.72);
      \node[tag] at (0,1.00) {a graphical \gls{iec}~\mbox{61131-3} \gls{ld} rung};
    \end{scope}

    \node[defn, text width=8.4cm] (kesbmc) at (0.0,1.5) {%
      \textbf{\kesbmc{}}\\an executable formal semantics of \gls{ld}, written as a set of K rewrite rules (extending \kst{}, the K semantics of \gls{st})
    };

    \draw[flow] (0,3.25) -- (kesbmc.north)
      node[tag, midway, right=1pt] {modelled, rung by rung};

    \node[tool, text width=4cm] (krun)   at (-2.9,-1.15) {%
      \code{krun}\, $\Rightarrow$\, \textbf{interpreter}\\
      {\footnotesize runs the diagram \emph{scan by scan}}};
    \node[tool, text width=4cm] (kprove) at (2.9,-1.15) {%
      \code{kprove}\, $\Rightarrow$\, \textbf{verifier}\\
      {\footnotesize machine-checked proofs}};

    \draw[flow] (kesbmc.south) -- (krun.north);
    \draw[flow] (kesbmc.south) -- (kprove.north);
    \node[tag, align=center, fill=none] at (0,-0.2)
      {one definition,\\both tools\\ \emph{for free}};

    \node[ext, text width=4.5cm] (plc) at (-2.9,-3.4) {%
      {\footnotesize reference \gls{plc} runtime (\openplc{}/\gls{matiec})}};
    \draw[<->, thick] (krun.south) -- (plc.north)
      node[tag, midway, align=left, fill=white]
      {scan-for-scan fidelity check};

    \node[ext, text width=4.5cm] (obl) at (2.9,-3.4) {%
      {\footnotesize properties discharged\\(e.g.\ scan determinism, safety)}};
    \draw[flow] (kprove.south) -- (obl.north);

  \end{tikzpicture}
  }

    \caption{\kesbmc{} in one picture. A graphical \gls{ld} rung is given an \emph{operational} meaning as a set of K rewrite rules -- extending \kst{}, the K semantics of \gls{st}, with contacts, coils, the scan cycle, timers, counters, and edge blocks. From that \emph{single} definition, K yields, for free, both an interpreter (\code{krun}) that \emph{runs} diagrams scan by scan and a program verifier (\code{kprove}). Because \kesbmc{} executes, its fidelity is checked directly, scan for scan, against a reference \gls{plc} runtime.}
    \Description{Flowchart illustrating the K-ESBMC approach for Ladder Diagram verification. Top: a graphical IEC 61131-3 LD rung with contacts (X0 normally-open, X1 normally-closed) and a coil (Y0). This rung feeds into the single K definition: K-ESBMC, an executable formal semantics of LD, written as K rewrite rules that extend K-ST for Structured Text. From this single definition, the K framework automatically generates two tools: krun (an interpreter that runs diagrams scan by scan) and kprove (a verifier for machine-checked proofs). The interpreter is validated through scan-for-scan fidelity checks against a reference PLC runtime (\openplc{}/\gls{matiec}). The verifier discharges properties such as scan determinism and safety.}
\label{fig:kesbmc-overview}
\end{figure}

%% file: fig_k_framework.tex
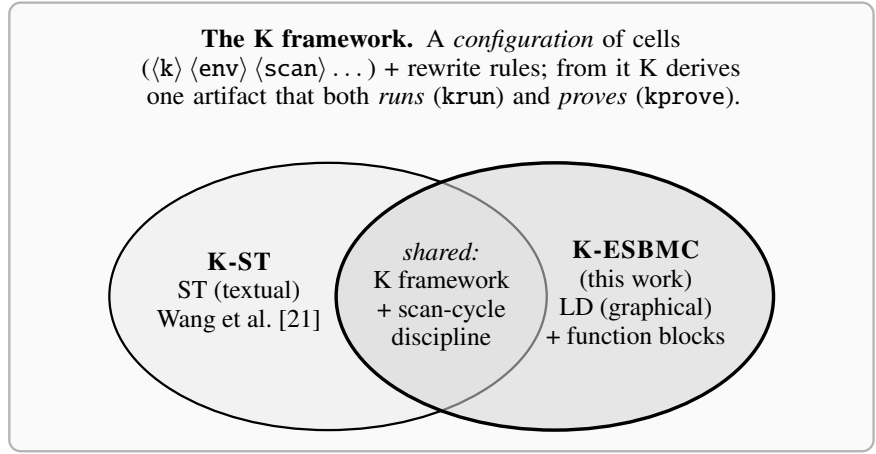
\begin{figure}[htbp]
    \centering
    \resizebox{0.7\linewidth}{!}{%
    \begin{tikzpicture}[font=\small]

    \node[align=center, text width=10cm] (hdr) at (0,2.65) {%
      \textbf{The K framework.} A \emph{configuration} of cells ($\langle\texttt{k}\rangle\,\langle\texttt{env}\rangle\,\langle\texttt{scan}\rangle\dots$) + rewrite rules; from it K derives one artifact that both \emph{runs} (\code{krun}) and \emph{proves} (\code{kprove}).\\[2pt]
      };

    \draw[thick, fill=black!7, fill opacity=0.55]
      (-1.4,-0.15) ellipse [x radius=2.7, y radius=1.65];
    \draw[very thick, fill=black!16, fill opacity=0.55]
      (1.4,-0.15) ellipse [x radius=2.7, y radius=1.65];

    \node[align=center, font=\footnotesize] at (-2.5,-0.1) {%
      \textbf{\kst{}}\\ \gls{st} (textual)\\ \citet{Wang2023}};

    \node[align=center, font=\footnotesize] at (2.4,-0.1) {%
      \textbf{\kesbmc{}}\\ (this work)\\ \gls{ld} (graphical)\\
      + function blocks};

    \node[align=center, font=\footnotesize, inner sep=1.5pt,
          fill=none, fill opacity=0.82, text opacity=1] at (0,-0.15) {%
      \emph{shared:}\\ K framework\\ + scan-cycle\\ discipline};

    \begin{scope}[on background layer]
      \node[draw, rounded corners, thick, black!30, fill=black!2,
            fit={(hdr) (-4.15,-1.85) (4.15,1.55)}, inner sep=6pt] {};
    \end{scope}

  \end{tikzpicture}
  }

    \caption{Where \kesbmc{} fits in K. K defines a language as a \emph{configuration} (cells) plus rewrite rules, deriving both an interpreter (\code{krun}) and a prover (\code{kprove}) from one definition. K covers C, Java, JavaScript, and the \gls{evm}. For \gls{iec}~61131-3, \citet{Wang2023} give \kst{} for \gls{st}; \kesbmc{} shares the framework and scan cycle but adds \gls{ld} and its function blocks, which \kst{} omits.}

    \Description{Venn diagram showing K-ESBMC's position relative to K-ST. The left lobe represents K-ST for Structured Text (textual) by Wang et al. (2023). The right lobe represents K-ESBMC (this work) for Ladder Diagram (graphical) plus function blocks. The overlapping region shows shared components: the K framework and scan-cycle discipline. The header describes the K framework's configuration of cells and rewrite rules, noting existing semantics for C, Java, JavaScript, and EVM applied to IEC 61131-3.}

  \label{fig:k-framework}
\end{figure}

%% file: fig_matiec_reference.tex
\begin{figure}[htbp]
  \centering
  \begin{tikzpicture}[
      font=\small,
      >={Latex[length=2mm]},
      box/.style  = {draw, rounded corners, align=center, inner sep=4pt, minimum height=9mm, fill=black!4},
      chip/.style = {draw, rounded corners=2pt, inner sep=3pt, font=\ttfamily\footnotesize, minimum width=8mm, fill=black!10},
      zoom/.style = {dashed, thin, black!45},
      tag/.style  = {font=\footnotesize\itshape, inner sep=2pt},
    ]

    \node[box, text width=2.3cm] (prog) at (-4.0,0)
      {\gls{iec}~\mbox{61131-3}\\program (\gls{st}/\gls{ld})};
    \node[box, text width=2.1cm] (matiec) at (0,0)
      {\gls{matiec}\\compiles to C};
    \node[box, text width=2.5cm] (run) at (4.1,0)
      {scan loop on deployed open-hardware \gls{plc}};

    \draw[->, thick] (prog) -- (matiec)
        node[tag, midway, above] {\openplc{}};
    \draw[->, thick] (matiec) -- (run)
      node[tag, midway, above] {executes};
    \draw[zoom] (matiec.south west) -- (-3.3,-1.35);
    \draw[zoom] (matiec.south east) -- ( 3.3,-1.35);

    \draw[very thick, rounded corners, fill=black!5]
      (-3.3,-3.55) rectangle (3.3,-1.35);

    \node[align=center] at (0,-1.72)
      {\textbf{The behavioral reference}\\[-1pt]
       {\footnotesize \gls{matiec}'s reference C function-block bodies}};

    \node[chip] (ton) at (-2.4,-2.55) {TON};
    \node[chip] (tof) at (-1.2,-2.55) {TOF};
    \node[chip] (tp)  at ( 0.0,-2.55) {TP};
    \node[chip] (ctu) at ( 1.2,-2.55) {CTU};
    \node[chip] (ctd) at ( 2.4,-2.55) {CTD};

    \node[tag] at (0,-3.18)
      {driven by a controlled notion of elapsed time ($\Delta t$)};


  \end{tikzpicture}

    \caption{\openplc{}/\gls{matiec} as behavioral reference. \openplc{} uses \gls{matiec}~\cite{deSousa2014} to compile \gls{iec}~61131-3 to C and runs it on a scan loop. \gls{matiec} supplies reference C implementations of standard function blocks (\code{TON}, \code{TOF}, \code{TP}, \code{CTU}, \code{CTD}) driven by elapsed time. As these run on deployed open-hardware \glspl{plc}, their per-scan behavior is the reference for validating \kesbmc{}'s blocks, and the tie-breaker when a verifier and \kesbmc{} disagree.}
    
    \Description{Flowchart showing \openplc{}/\gls{matiec} as the behavioral reference. Top: IEC 61131-3 program (ST/LD) goes through the \gls{matiec} compiler to C, then executes as a scan loop on a deployed open-hardware PLC. A magnified callout box reveals \gls{matiec}'s reference C function-block bodies: TON, TOF, TP, CTU, and CTD, driven by a controlled notion of elapsed time (Δt). These reference implementations serve as the per-scan oracle for K-ESBMC's function blocks, and as the tie-breaker when a verifier and K-ESBMC disagree.}
  \label{fig:matiec-reference}
\end{figure}
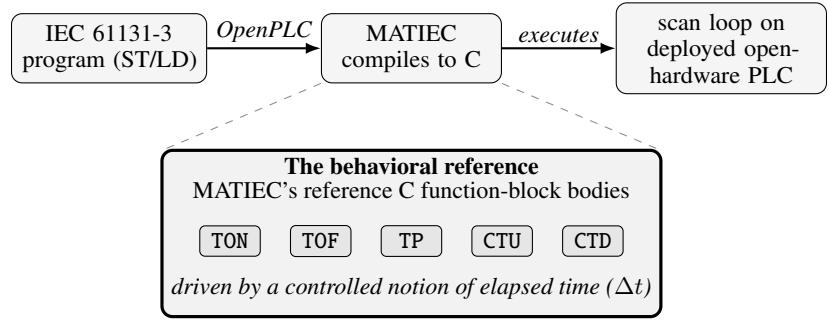

%% file: fig_positioning.tex
\begin{figure}[htbp]
  \centering
  \begin{tikzpicture}[
      font=\small,
      >={Latex[length=2mm]},
      box/.style     = {draw, rounded corners, align=center, inner sep=4pt, minimum height=8mm, fill=black!4},
      hl/.style      = {box, very thick, fill=black!8},
      callout/.style = {draw, dashed, rounded corners, align=center, inner sep=4pt, thick},
      tick/.style    = {thin, black!55},
      tag/.style     = {font=\footnotesize\itshape, inner sep=2pt, align=center},
    ]

    \draw[<->, thick] (-5.3,1.0) -- (5.3,1.0);
    \node[font=\footnotesize\scshape, anchor=south east] at (-2.9,1.08) {testing};
    \node[font=\footnotesize\scshape, anchor=south west] at ( 2.9,1.08) {verification};
    \node[tag, anchor=north east] at (-3.0,0.9) {empirical, lightweight,\\high coverage};
    \node[tag, anchor=north west] at (3.0,0.9) {mechanized, heavyweight,\\complete};

    \node[box, text width=4.0cm] (dt) at (-2.8,2.35)
      {\textbf{Differential\\ testing}\\{\scriptsize \citet{Yang2011csmith}}};
    \node[box, text width=4.0cm] (vc) at (2.8,2.35)
      {\textbf{Verified compilation \&\\translation validation}\\ {\scriptsize \citet{Pnueli1998translation, Leroy2009compcert}}};
    \draw[tick] (dt.south) -- (-2.8,1.0);
    \draw[tick] (vc.south) -- (2.8,1.0);

    \fill (0,1.0) circle (1.6pt);                 
    \draw[tick] (0,1.0) -- (0,-0.05);          

    \node[hl, text width=7.2cm] (kesbmc) at (0.0,-0.9) {%
      \textbf{\kesbmc{} (this work)}\\
      differential testing against an \emph{executable} K reference\\
      \textbf{+} machine-checked per-construct lemmas\\
      (enabled by K's dual \code{krun}\,/\,\code{kprove} generation)};

    \node[callout, text width=7.2cm] (new) at (0.0,-2.9) {%
      \textbf{New:} the reference-semantics method turned on a
      \emph{verifier's front-end}, in the cyber-physical \gls{plc}
      setting -- where the discrepancies are \emph{safety-relevant}};
    \draw[->, thick] (kesbmc) -- (new);

  \end{tikzpicture}

    \caption{Positioning on the assurance/effort spectrum. Rather than full verified compilation~\cite{Leroy2009compcert} or translation validation~\cite{Pnueli1998translation}, we adopt differential testing~\cite{Yang2011csmith} against an executable reference, supported by machine-checked per-construct lemmas. K's dual generation of interpreter and prover from a single semantics -- as with C~\cite{Ellison2012c} and the \gls{evm}~\cite{Hildenbrandt2018kevm} -- enables both testing and proving. Our contribution is adapting this approach to a \emph{verifier's} front-end in a safety-critical domain.}

    \Description{Positioning on the assurance/effort spectrum. A horizontal axis shows testing (empirical, lightweight, high coverage) on the left and verification (mechanized, heavyweight, complete) on the right. Two established paradigms are shown above the axis: Differential testing (Yang et al. 2011) on the left, and Verified compilation & translation validation (Pnueli et al. 1998, Leroy 2009) on the right. K-ESBMC (this work) is positioned in the middle, combining differential testing against an executable K reference with machine-checked per-construct lemmas, enabled by K's dual krun/kprove generation. A callout box highlights the novelty: applying the reference-semantics method to a verifier's front-end in the cyber-physical PLC setting, where discrepancies are safety-relevant.}
  \label{fig:positioning}
\end{figure}
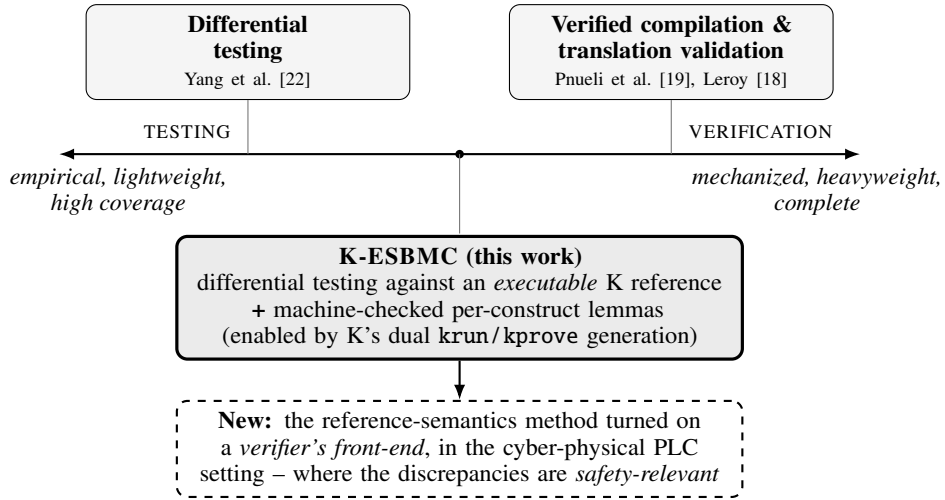

%% file: fig_arch.tex
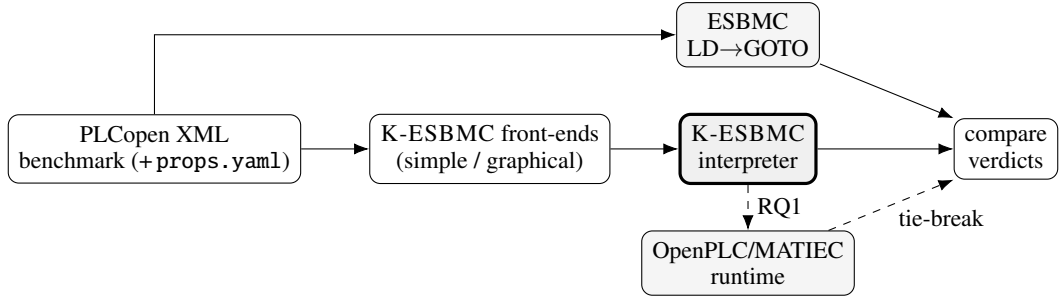
\begin{figure}[htbp]
  \centering
  \begin{tikzpicture}[
      font=\small,
      box/.style={draw, rounded corners, align=center, inner sep=4pt, minimum height=8mm},
      engine/.style={box, fill=black!4},
      oracle/.style={box, very thick, fill=black!6},
      >={Latex[length=2mm]},
      node distance=6mm and 9mm]
    \node[box] (xml) {PLCopen XML\\benchmark {\footnotesize(+\,\code{props.yaml})}};
    \node[box, right=of xml] (fe) {\kesbmc{} front-ends\\{\footnotesize(simple / graphical)}};
    \node[oracle, right=of fe] (kesbmc) {\kesbmc{}\\interpreter};
    \node[engine, above=of kesbmc] (esbmc) {\gls{esbmc}\\\gls{ld}$\rightarrow$GOTO};
    \node[engine, below=of kesbmc] (openplc) {\openplc{}/\gls{matiec}\\runtime};
    \node[box, right=18mm of kesbmc] (cmp) {compare\\verdicts};
    \draw[->] (xml) -- (fe);
    \draw[->] (fe) -- (kesbmc);
    \draw[->] (xml.north) |- (esbmc.west);
    \draw[->] (esbmc) -- (cmp.north west);
    \draw[->] (kesbmc) -- (cmp);
    \draw[->, dashed] (openplc) -- node[below, pos=0.6, font=\footnotesize] {\hspace{1cm}tie-break} (cmp.south west);
    \draw[->, dashed] (kesbmc) -- (openplc) node[midway, right, font=\footnotesize] {RQ1};
  \end{tikzpicture}
  \caption{The three-engine differential. Each PLCopen benchmark is verified by \gls{esbmc} (the translation under test) and, independently, executed by the \kesbmc{} oracle via frontends that never reuse \gls{esbmc}'s parser; verdicts are compared. On any disagreement, the \openplc{}/\gls{matiec} runtime -- against which \kesbmc{} is validated scan-for-scan (RQ1) -- is the external tie-breaker.}
  \Description{Three-engine differential verification architecture: PLCopen XML benchmarks with properties feed into K-ESBMC frontends (simple/graphical), which drive the K-ESBMC interpreter oracle. ESBMC (LD→GOTO translation under test) and \openplc{}/\ runtime receive separate inputs. All three engines feed into a comparator that checks verdict agreement, with \openplc{}/\gls{matiec} serving as the external tie-breaker on disagreements.}
  \label{fig:arch}
\end{figure}

%% file: 0.ack.tex
\section*{Acknowledgments}
The authors would like to express their gratitude to the Department of Computer Science at the University of Manchester (UoM) and the Systems and Software Security (S3) Research Group for their invaluable support, collaborative environment, and access to cutting-edge resources, which were instrumental in the success of this research. We conducted this work with partial funding from the Engineering and Physical Sciences Research Council (EPSRC) grants EP/T026995/1, EP/V000497/1, EP/X037290/1, and the Soteria project, awarded by the UK Research and Innovation under the Digital Security by Design (DSbD) Programme.